\documentclass[manuscript]{acmart}

\usepackage{mathtools}
\usepackage{subcaption}
\usepackage{graphicx}
\usepackage{array}
\usepackage{makecell}

\AtBeginDocument{%
  }

\setcopyright{acmlicensed}
\copyrightyear{2018}
\acmYear{2018}
\acmDOI{XXXXXXX.XXXXXXX}

\acmConference[Conference acronym 'XX]{Make sure to enter the correct
  conference title from your rights confirmation emai}{June 03--05,
  2018}{Woodstock, NY}
\acmISBN{978-1-4503-XXXX-X/18/06}

\begin{document}

\title{Beyond Instructed Tasks: Recognizing In-the-Wild Reading Behaviors in the Classroom Using Eye Tracking}

\author{Eduardo Davalos}
\email{eduardo.davalos.anaya@vanderbilt.edu}
\orcid{0000-0001-7190-7273}
\affiliation{%
  \institution{Vanderbilt University}
  \city{Nashville}
  \state{TN}
  \country{USA}
}

\author{Jorge Alberto Salas}
\email{jorge.a.salas@vanderbilt.edu}
\orcid{ 0000-0002-8885-0813 }
\affiliation{%
  \institution{Vanderbilt University}
  \city{Nashville}
  \state{TN}
  \country{USA}
}

\author{Yike Zhang}
\email{yike.zhang@vanderbilt.edu}
\orcid{0000-0003-3503-2996}
\affiliation{%
  \institution{Vanderbilt University}
  \city{Nashville}
  \state{TN}
  \country{USA}
}

\author{Namrata Srivastava}
\email{namrata.srivastava@vanderbilt.edu}
\orcid{0000-0003-4194-318X}
\affiliation{%
  \institution{Vanderbilt University}
  \city{Nashville}
  \state{TN}
  \country{USA}
}

\author{Yashvitha Thatigotla}
\email{yashvitha.thatigotla@vanderbilt.edu}
\orcid{0009-0005-2035-752X}
\affiliation{%
  \institution{Vanderbilt University}
  \city{Nashville}
  \state{TN}
  \country{USA}
}

\author{Abbey Gonzales}
\email{abbey.e.gonzales@vanderbilt.edu}
\orcid{0009-0007-7374-8852}
\affiliation{%
  \institution{Vanderbilt University}
  \city{Nashville}
  \state{TN}
  \country{USA}
}

\author{Sara McFadden}
\email{sara.mcfadden@vanderbilt.edu}
\orcid{0000-0001-6375-1272}
\affiliation{%
  \institution{Vanderbilt University}
  \city{Nashville}
  \state{TN}
  \country{USA}
}

\author{Sun-Joo Cho}
\email{sj.cho@vanderbilt.edu}
\orcid{0000-0002-2600-8305}
\affiliation{%
  \institution{Vanderbilt University}
  \city{Nashville}
  \state{TN}
  \country{USA}
}

\author{Gautam Biswas}
\email{gautam.biswas@vanderbilt.edu}
\orcid{0000-0002-2752-3878}
\affiliation{
  \institution{Vanderbilt University}
  \city{Nashville}
  \state{TN}
  \country{USA}
}

\author{Amanda Goodwin}
\email{amanda.goodwin@vanderbilt.edu}
\orcid{0000-0002-6439-7399}
\affiliation{%
  \institution{Vanderbilt University}
  \city{Nashville}
  \state{TN}
  \country{USA}
}

\renewcommand{\shortauthors}{Davalos et al.}

\begin{abstract}





Understanding reader behaviors such as skimming, deep reading, and scanning is essential for improving educational instruction. While prior eye-tracking studies have trained models to recognize reading behaviors, they often rely on instructed reading tasks, which can alter natural behaviors and limit the applicability of these findings to in-the-wild settings. Additionally, there is a lack of clear definitions for reading behavior archetypes in the literature. We conducted a classroom study to address these issues by collecting instructed and in-the-wild reading data. We developed a mixed-method framework, including a human-driven theoretical model, statistical analyses, and an AI classifier, to differentiate reading behaviors based on their velocity, density, and sequentiality. Our lightweight 2D CNN achieved an F1 score of 0.8 for behavior recognition, providing a robust approach for understanding in-the-wild reading. This work advances our ability to provide detailed behavioral insights to educators, supporting more targeted and effective assessment and instruction.


\end{abstract}


  
\begin{teaserfigure}
  \centering
  \includegraphics[width=0.9\textwidth,trim={0 0.5cm 0 3.25cm},clip]{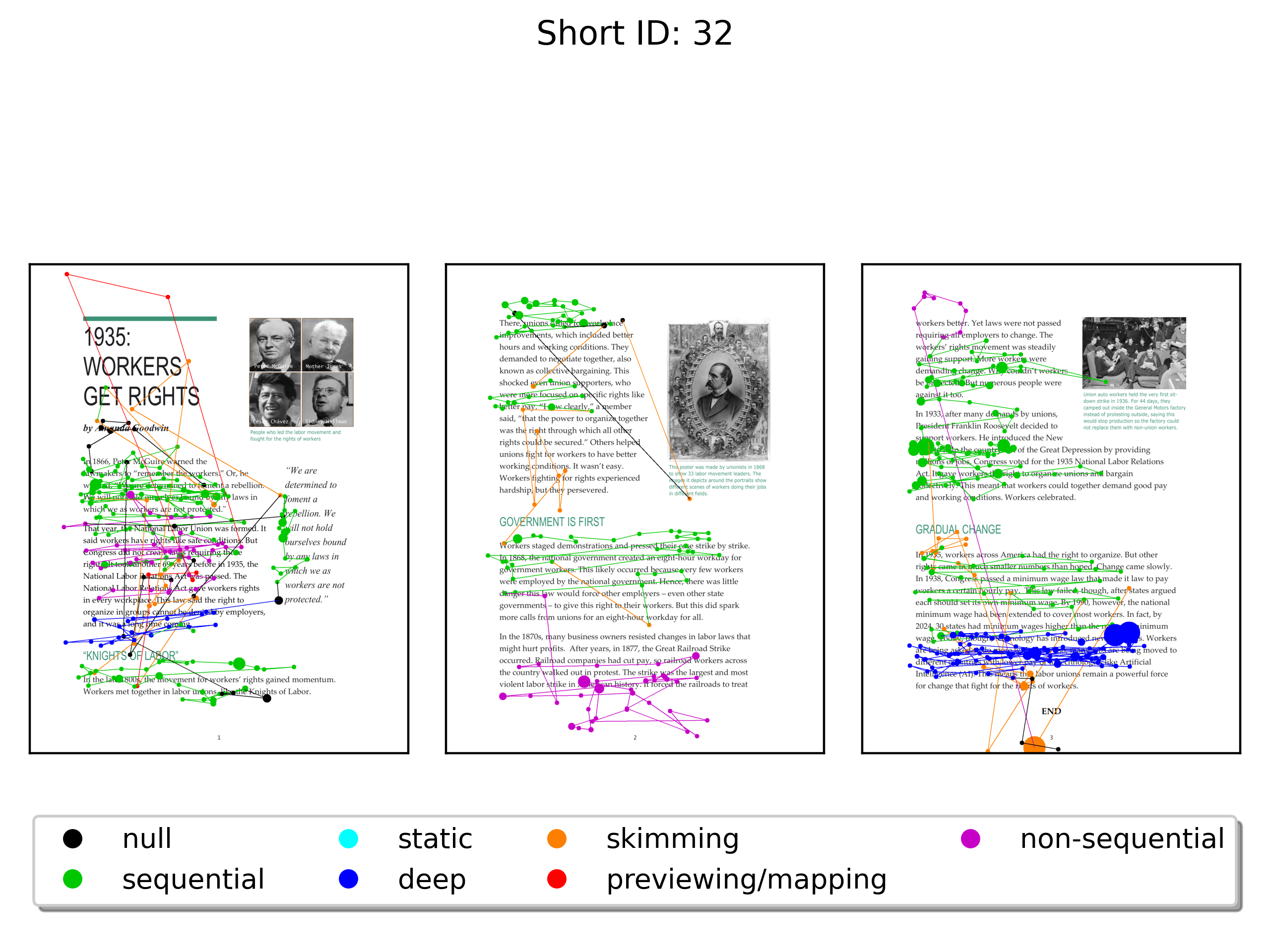}
  \Description{Example of human-labeled In-the-Wild reading behaviors overlaid on to a PDF}
  \caption{\textbf{In-the-Wild Reading Behaviors in Naturalistic Hierarchical Passages}: Displaying the human-labeled reading behavior annotations onto an authentic reading passage.}
  \Description{}
  \label{fig:teaser}
\end{teaserfigure}

\received{20 February 2007}
\received[revised]{12 March 2009}
\received[accepted]{5 June 2009}

\maketitle


\section{Introduction}


Eye-tracking technology has become an invaluable tool for analyzing reading processes and understanding how readers comprehend text \cite{Just1980AComprehension, Meziere2023UsingComprehension, Meziere2024ScanpathComprehension, Southwell2020WhatTexts, Caruso2022, Pena2024EyeTracking}. Researchers can collect detailed data on gaze behaviors by tracking eye movements, such as fixation duration, saccade length, and regressions \cite{Southwell2020WhatTexts}. These accumulative gaze metrics have been used to draw correlations between reading strategies and learning outcomes \cite{RajendranPredictingBehavior}, shedding light on the cognitive processes involved in text comprehension \cite{Meziere2023UsingComprehension, Caruso2022, Meziere2024ScanpathComprehension, Southwell2020WhatTexts}. For example, higher fixation counts and longer fixation durations may indicate deeper cognitive processing \cite{Negi2020FixationVideos}, while shorter and more scattered fixations could suggest skimming or scanning behaviors \cite{Kelton2019ReadingReal-time, Chen2023}. While these quantitative metrics provide a wealth of data on how individuals interact with text, they often lack the interpretive depth to translate findings into practical applications in educational settings.

Despite the wealth of data provided by eye-tracking studies, a significant limitation of the analyses conducted is that the gaze metrics are challenging to interpret and do not offer actionable insights for educators. While researchers can identify patterns that correlate with comprehension or engagement \cite{Lee2021WhenModels, Miller2015UsingEngagement, Ahn2020}, these findings rarely translate into clear guidelines or strategies for teachers to implement in the classroom. There is often a disconnect between what the data reveals about reading processes and how educators can use that information to support and improve student learning. Without clear feedback mechanisms, teachers may struggle to adjust their instruction based on eye-tracking data, limiting the potential impact of this research on educational practice.


To address these limitations, other approaches in eye-tracking research have focused on recognizing specific reader behaviors, such as ``skimming'', ``deep reading'', and ``scanning''. Identifying these distinct behaviors during reading is a crucial first step in providing meaningful, context-specific feedback to educators, enabling them to understand better and respond to students' reading strategies. Many existing methods have successfully classified behaviors like skimming versus deep reading \cite{Chen2023, Biedert2012}, showing that it is possible to infer reading strategies from gaze data. However, a major limitation of these studies is their reliance on task instructions to induce specific reading behaviors, such as asking participants to skim or read deeply. This approach simplifies the data labeling process by ensuring that each segment of gaze data is associated with a known behavior, minimizing the need for costly human annotation.

However, providing specific instructions to participants on how to read can significantly alter their natural reading behaviors \cite{Jian2018ReadingStudy}, impacting the validity of the findings when applied to real-world, in-the-wild settings. Reading is often more fluid and context-dependent in authentic environments, with readers dynamically switching between behaviors based on their comprehension needs, interest levels, or the task at hand \cite{Jarodzka2017TrackingReading}. The prescriptive nature of task-based reading studies can limit the generalizability of these models and frameworks to everyday reading situations, where behavior is not fixed but adaptive. To the best of our knowledge, no studies have specifically analyzed the differences in reading behaviors between instructed and naturalistic settings and their implications, revealing a critical gap in the literature that must be addressed to develop more ecologically valid insights into reading processes.

Therefore, in this study, we address the limitations of prior research on reading behaviors by performing an exploratory investigation on both instructed and in-the-wild reading conditions. We conducted a classroom-based study that collected eye-tracking data from students engaged in natural and instructed reading tasks. Our findings reveal significant differences between instructed and in-the-wild reading behaviors, particularly in the variability and dynamics of gaze patterns, which are crucial for understanding authentic reading processes. We developed a qualitative theoretical framework and conducted statistical evaluations to classify reading behaviors. Next, we developed a real-time classifier using a lightweight 2D convolutional neural network (CNN) that outperformed traditional models for detecting reading behaviors. These insights provide a deeper understanding of natural reading behaviors and highlight the importance of incorporating in-the-wild data for developing reliable reading behavior recognition systems in educational contexts. Overall, in this paper, we present several key contributions that advance the field of reading behavior research:

\begin{itemize}
    \item \textbf{Datasets}: We introduce the first publicly available datasets capturing gaze behavior for both instructed and in-the-wild reading contexts. This novel dataset fills a critical gap in the field by enabling further research into naturalistic reading behaviors.
    \item \textbf{Exploratory Framework and Taxonomy for Reading Behaviors}: Building on qualitative human coding, we introduce a proof-of-concept framework and taxonomy for classifying reading behaviors. This exploratory contribution aims to establish foundational definitions and classifications, serving as a starting point for future research to refine and expand upon in the reading activity research community.
    \item \textbf{Feasibility of Real-Time Behavior Classification}: We present a 2D neural network-based real-time behavior classifier as a proof-of-concept for in-the-wild reading behavior classification. Rather than aiming for broad generalizability, this model demonstrates the feasibility of using AI to recognize complex reading behaviors in real-time, addressing a significant challenge in current behavioral research.

\end{itemize}



We will make our code and datasets publicly available on GitHub at \textcolor{red}{<Link>} to promote open science research and enhance collaboration in understanding reader behaviors through a comprehensive framework, taxonomy, and classifier.

\section{Related Work}\label{sec:background}

This related work section reviews the application of eye-tracking in reading comprehension to understand gaze patterns and cognitive processes, highlights approaches to reading behavior detection using both rule-based and machine learning methods, and identifies literature gaps such as inconsistent behavior taxonomies, limited ecological validity in studies, and challenges in recognizing diverse, naturalistic reading behaviors.

\subsection{Reading Comprehension and Eye-Tracking}
Applying eye-tracking in reading comprehension and analyzing the reading process has a long and vast tradition for understanding the underlying gaze patterns that partake in reading. \citet{Just1980AComprehension} applied eye-tracking to understand reading processes by analyzing eye fixation patterns to explore how readers allocate visual attention during reading. Their studies demonstrated that eye movements, such as gaze duration on words and fixations on specific text elements, provide insight into the cognitive processes involved in word recognition, comprehension, and integration of information across sentences. Various aspects of reading have been explored through the use of gaze data, such as reader subtyping \cite{Ma2023FromReading}, reading comprehension prediction \cite{Meziere2023UsingComprehension, Meziere2024ScanpathComprehension, Caruso2022, Southwell2020WhatTexts}, reading behavior recognition \cite{Chen2023, RajendranPredictingBehavior, Davalos2023IdentifyingGraphs}, and cognitive load prediction \cite{Zu2018UseTheory}. In many of these studies, a reading stimulus is presented to the participant with a delegated task, such as answering questions, retrieving information, or summarizing the text. While the participant is going through the experiment, eye-tracking time-series data is collected. Fixations, where the eyes remain focused on a single point, and saccades, the rapid eye movements between fixations, are key metrics in eye-tracking research. They provide insight into where and how long a reader focuses on specific parts of a text, helping researchers understand cognitive processes like comprehension, reading ability, and behavior. These eye movements are estimated using filtering algorithms and are analyzed further by generating accumulative or down-sampled representations that can be linked to participant variables.

Through this study structure, \citet{Meziere2024ScanpathComprehension} investigated the potential of eye-tracking measures as a tool for assessing reading comprehension by examining the relationship between eye movements and comprehension scores across three widely used reading tests. Their findings suggested that while eye-tracking measures could predict reading comprehension, the effectiveness of specific measures varied depending on the test format and task demands, underscoring the complex nature of reading comprehension assessments. Complementary to their findings, reading comprehension measures collected from standard reading tests differ from naturalistic reading \cite{Kaakinen2010TaskReading}. This is because authentic reading involves the composition of multiple skills and techniques, with natural reading having a broader range of tasks -- long readings, short readings, different purposes, and different formats -- compared to the narrow and limited standardized reading tests. Moreover, the accumulative eye-tracking metrics used to analyze reading comprehension lose the natural temporal progression and transition, lowering the resolution of our analysis and understanding of the innate complexity of reading and gaze data.

\subsection{Reading Behavior Detection}
In terms of activity recognition for reading behaviors, \citet{Campbell2001} utilized a rule-based algorithm to distinguish between ``reading'' and ``skimming'' by analyzing eye movements such as saccades and fixations. Their approach involved assigning specific point values to different eye movement patterns to detect when users were engaged in reading versus other types of visual scanning. Instead of a rules-based approach, \citet{Kelton2019ReadingReal-time} introduced a method for real-time reading detection using a Region Ranking Support Vector Machine, identifying ``reading'' vs. ``skimming". Their approach combined both global and local information to accurately classify reading versus skimming behaviors and achieve up to 82.5\% accuracy in predicting reading behavior. Other methods \cite{Srivastava2018CombiningRecognition, BektaGEAR_MANUAL, Chen2023} employ a time-window approach, where sliding time-window segments the gaze data and accumulative metrics are computed based on the isolated segment. More specifically, \citet{Chen2023} employed the time-window technique to classify reading behaviors as ``deep'' or ``skimming''. These time-window methods have achieved good performance across various studies and eye-tracking hardware. However, a time-window approach is sensitive and generally have lower performance when using smaller time windows. To achieve robust behavior classification, time-window approaches require a larger time window (>30 seconds) but this limits the level of granularity in models' behavior recognition. Lastly, it is still uncertain which reading behaviors should be recognized, as previous studies use varying behavior labels and definitions. Moreover, many of these definitions are purpose-driven. For instance, \citet{Campbell2001} distinguishes ``scanning'' from ``skimming'' based on their objective: ``scanning'' is typically defined as a targeted ``skim'' to locate a specific detail in the text, while ``skimming'' is a more general strategy for quickly navigating through the text to gain a surface-level understanding of its content and layout. These definitions are logically grounded, particularly since many derive from traditional ``pen and paper'' studies, which rely on manually logging metrics and behaviors without eye-tracking, where the reading purpose is predetermined. However, when considering only the gaze scanpath data, differentiating between ``skimming'' and ``scanning" becomes ambiguous, as the eye movements in both behaviors appear similar.

\subsection{Literature Gaps and Limitations} Through our review of the relevant literature, we have identified critical gaps regarding readi recognition. These include the following: \begin{itemize}
    \item \textbf{Authentic vs. Instructed Behavior}: Current reader behavior activity literature instructs students on how to read (e.g., skim, deeply read). However, naturalistic reading might not align with these task-instructed behaviors and limits the applicability of these methods.
    \item \textbf{Inconsistent taxonomy}: The lack of precise  definitions of reading behaviors with illustrative examples limits the research community's ability to effectively communicate, discuss, and share resources and datasets to achieve reliable and robust reading behavior detection. 
    \item \textbf{Lack of Ecological Studies}: Many gaze-based reading studies are performed in laboratory settings and use unauthentic texts such as homogeneous large spacing single paragraphs instead of typical hierarchical text with titles, headings, captions, and paragraphs. These study constraints can cause to alter the reading behavior exhibited by participants.
    \item \textbf{Behavior Classification Granularity}: Prior time-window approaches limit the resolution of behavior prediction, likely missing brief yet critical reading behaviors like skimming.
\end{itemize}

\subsection{Research Questions}


To address the gaps in the literature, we formulate the following research questions:

\begin{itemize}

    \item \textbf{RQ1}: What are the reading behaviors in naturalistic and instructed reading conditions? Do the behaviors differ and/or do their characteristics differ?
    \item \textbf{RQ2}: How can we develop a theoretical framework for reading behaviors -- comprising behavior labels and definitions -- based on human observations while ensuring that reading behaviors are measurable and distinguishable using data-driven methods?
    \item \textbf{RQ3}: How can we effectively classify reading behaviors, including high-speed and transitional behaviors?
\end{itemize}

\section{Study Design}\label{sec:study_design}

\begin{figure}[h]
    \centering
    \includegraphics[width=1\linewidth]{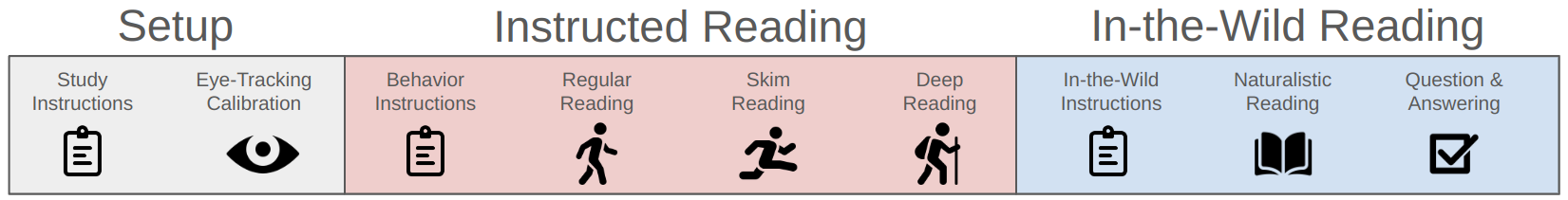}
    \Description{Flow diagram of the study structure. 3 sequential blocks with the titles: setup, instructed behavior, in-the-wild reading. Within the setup, we provide study instructions and perform eye-tracking calibration. In the instructed reading block, we have behavior instructions, regular reading, skim reading, and deep reading sub-activities. In the In-the-Wild reading block, we provide instructions, reading passage, and a set of questions.}
    \caption{\textbf{Study Structure}: The study is organized into three main sections: setup, instructed behavior activity, and in-the-wild reading. During the setup phase, students completed an eye-tracking calibration and received instructions. In the instructed behavior activity, students were directed to read in a specified manner. For the in-the-wild reading section, students were asked to read a passage as they would naturally in a classroom setting, without access to any questions. After completing the reading, they moved on to the question segment, where they could access both the questions and the text. However, they could not revisit previous questions once they had submitted an answer for each one.}
    \label{fig:study_design}
\end{figure}

The study involved an ecological setting with 38 sixth-grade students to examine natural reading behaviors using eye-tracking technology. An overview of the study is shown in Fig. \ref{fig:study_design}. Students were asked to perform two tasks: (1) instructed behavior reading, where they followed specific reading strategies like skimming or deep reading, and (2) uninstructed reading of a passage, consisting of a ``coldread'' segment without questions and a ``questions'' segment with text access. Gaze data was collected using Tobii Pro Spark\footnote{\href{https://nbtltd.com/wp-content//uploads/2022/12/tobii_pro_spark_product_description.pdf}{Tobii Pro Spark Specification Sheet PDF}} eye-trackers and a custom web application interface, allowing for detailed tracking and analysis of reading patterns. Preprocessing was conducted to align gaze data with the text, enabling accurate observation of reading behaviors. The analysis focused primarily on the coldread segment to maintain consistency in examining natural reading processes. This study was approved by [Anonymous] University's Ethics Committee.

\subsection{Participants}

We conducted a classroom study with 38 sixth-grade students from a private school in the southeastern region of the United States. Based on the demographic data provided by the teacher, 22 students identified as male, and 16 students identified as female. The racial composition was predominately White (\textit{N}=35), with 2 Black and 1 Asian student. Among the White students, 2 identified as Hispanic or Latino. From the 38 participants' data, we ruled out 11 students based on the following criteria: 2 students did not follow study protocols, 4 students' data was corrupted (impacted by students' covering their faces with their hands), and 5 students' data were too noisy for human labeling (impacted by posture). A total of N=27 participant sessions were used for our analysis.

\subsection{Stimulus, Apparatus, and Procedure}

\begin{figure}[!ht]
    \centering
    \includegraphics[width=0.8\linewidth]{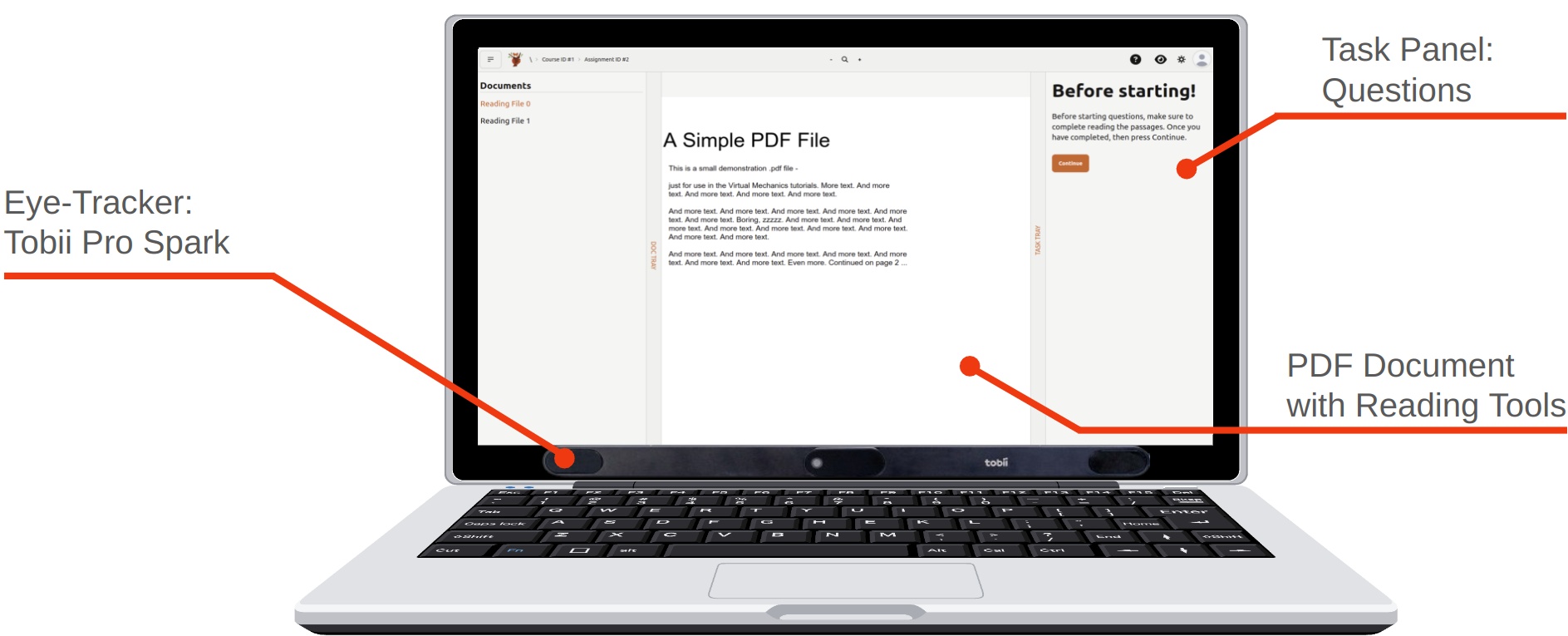}
    \Description{Image of HP computer with the custom web application. The placement of the Tobii Pro Spark eye-tracking device is indicated at the bottom of the display.}
    \caption{\textbf{Data Collection Protocol}: Deployed using consumer-level HP laptops to present the custom web application and a Tobii Pro Spark to collect eye-tracking data. The web application provides a digital PDF reader with a side panel for questions.}
    \label{fig:protocol}
\end{figure}

In the study, students performed two tasks: (1) an instructed behavior task, where they read sections of a PDF with specific instructions -- such as ``skim'', ``regularly'' read, or ``deeply'' read -- for about 20 seconds to minimize cross-contamination of behaviors \cite{Rayner2012PsychologyReading}; and (2) an in-the-wild task, where they read a PDF passage naturally, as they would in class, followed by a question-answering session with access to the text. The passage was approximately 500 words, aligned with the 6th-grade Common Core Standards, with a Lexile level of 810-1000. Prior to reading, students underwent a 9-point calibration with a Tobii Pro Spark eye-tracker (60 Hz). During the session, gaze data and digital interaction logs were collected via a custom web application and ETProWeb, leveraging the Tobii Pro SDK to integrate gaze data directly within the web interface. The study used HP 15-dy2xxx laptops with a resolution of 1920x1080 pixels for data collection. The complete protocol is shown in Fig. \ref{fig:protocol}.

For the analysis presented in this paper, we focused on exploring the reading behaviors in the instructed task and the coldread section of the in-the-wild task. The aim is to understand the reading process and the behaviors exhibited by students. In the in-the-wild subsection where students perform two tasks -- coldread and question-answering --, we identified a notable shift in behaviors. By asking students to answer questions with the text accessible, their behaviors and strategies drastically changed from reading-focused to information-retrieval. Given the different task objectives and our focus is on examining reading behaviors between instructed and in-the-wild conditions, we isolated our in-the-wild corpus to the cold reading task to concentrate our analysis on reading-focused behaviors.

\subsection{Preprocessing}


With the recorded gaze data, our goal was to analyze the gaze scanpath on the PDF content to understand reading behaviors and how students approached text comprehension. For this, it was essential to have the gaze data in page coordinate space (PCS), ensuring consistent and reliable analysis of gaze patterns on the text's pages. Traditional screen-based eye-tracking methods, however, record gaze data in screen coordinate space (SCS), where the $(0,0)$ origin is positioned at the top-left corner of the screen. This approach does not account for dynamic movements, such as scrolling and zooming, that occur in a web-based PDF viewer.

\begin{figure}[!ht]
    \centering
    \includegraphics[width=0.8\linewidth]{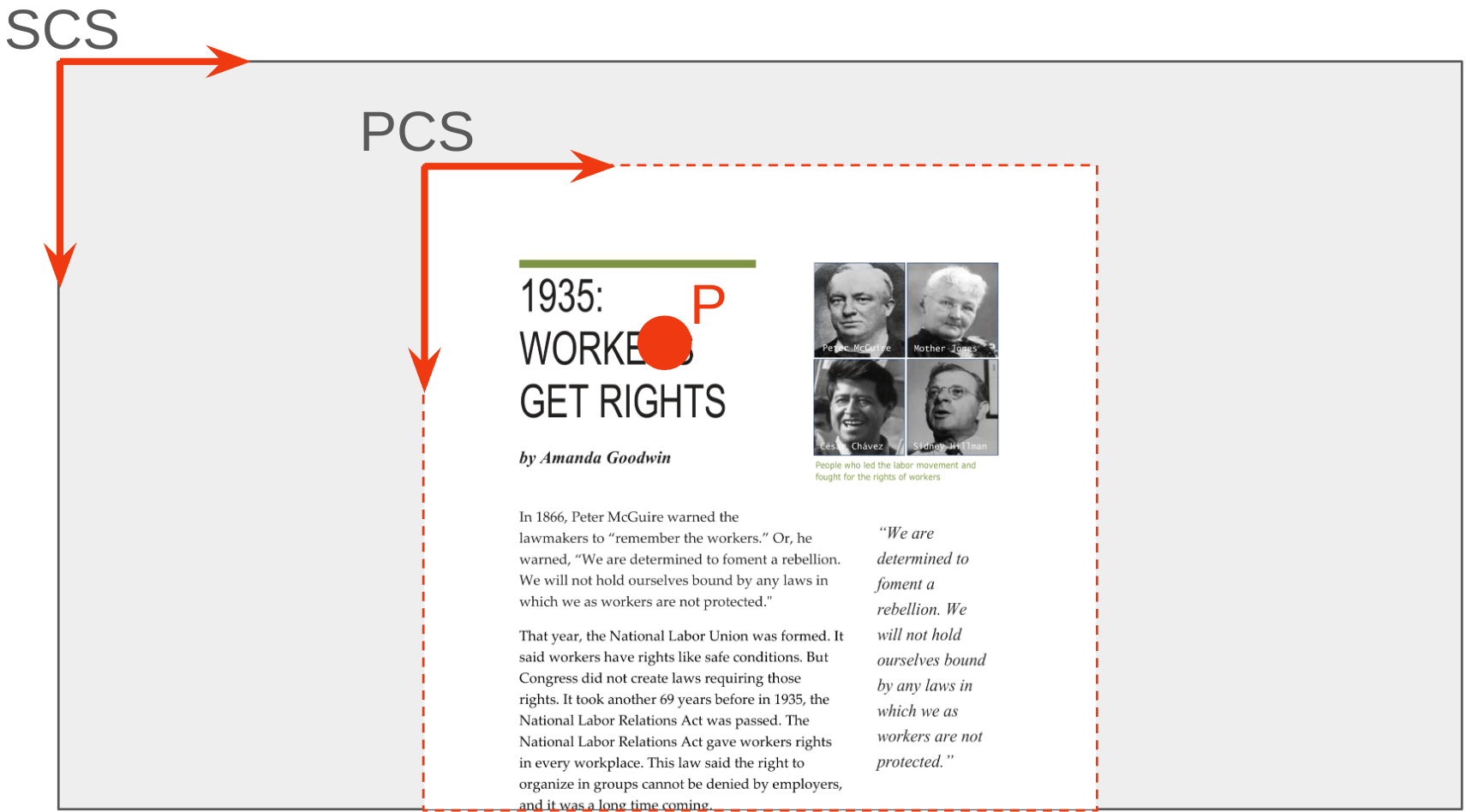}
    \Description{Illustration of the relationship between screen coordinate space (SCS) and an example page coordinate space (PCS) of the PDF viewer within the web application.}
    \caption{\textbf{Page AOI Encoding:} Illustration of the relationship between screen coordinate space (SCS) and an example page coordinate space (PCS) of the PDF viewer within the custom web application.}
    \label{fig:gaze_preprocessing}
\end{figure}

To address this, we performed real-time area-of-interest (AOI) encoding by projecting the gaze onto the text, as shown in Fig. \ref{fig:gaze_preprocessing}. The PDF's pages are tracked by the document object model (DOM) and new incoming $P_{scs}$ gaze points are transformed to $P_{pcs}$ using the following equation:

\begin{equation}
    P_{pcs} = \left(\frac{P_{scs}(0) - R_l}{R_w}, \frac{P_{scs}(1) - R_t}{R_h}\right),
\end{equation}
where $R$ is the bounding rectangle of the page element in the DOM, with access to its left-most x value $l$, top-most y value $t$, width $w$, and height $h$. Through this coordinate conversion for the gaze, we perform our gaze scanpath analysis based on the PDF content. Additionally, analyzing the gaze within PCS makes our findings independent of screen resolution.

\begin{figure}[h]
    \centering
    \includegraphics[width=\linewidth, trim={0cm 2.5cm 0cm 3.25cm}, clip]{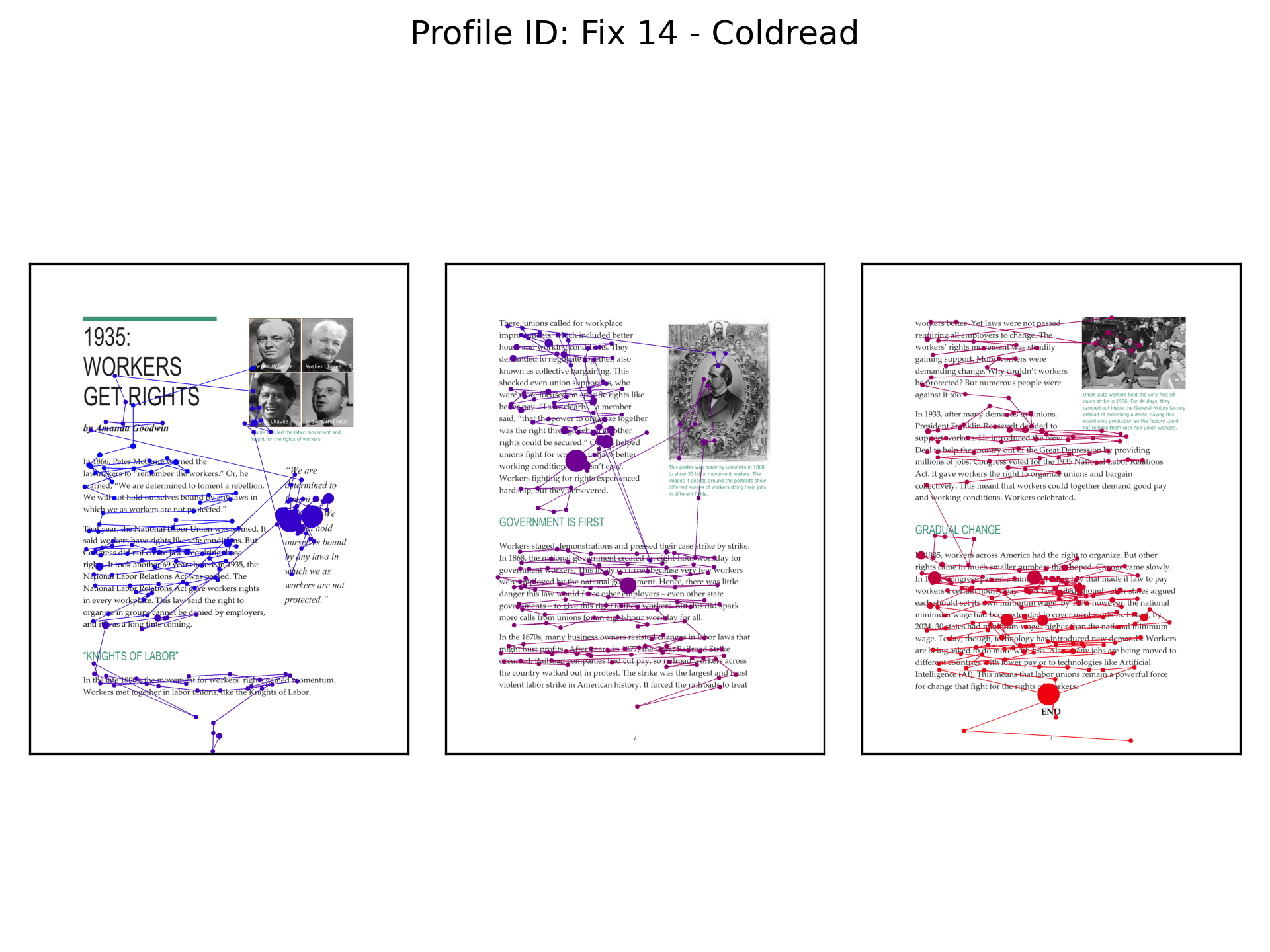}
    \Description{Example of a gaze scanpath that has been colored with a blue-to-red gradient that has been overlaid on top of the PDF text.}
    \caption{\textbf{Example gaze scanpath:} Utilizing a blue-to-red gradient to illustrate the temporal evaluation of the scanpath. The visualization is overlaid on top of the text to help contextualize the gaze. The beginning of the scanpath is painted blue $(0,0,255)$, throughout time the blue is replaced with red, and at the end is marked with full red color $(255,0,0)$.}
    \label{fig:gaze_scanpath}
\end{figure}

With the gaze data successfully converted to PCS, we could create visualizations and compute metrics to observe, compare, and interpret reading behaviors. Previous methods involved overlaying gaze scanpaths on the text to visualize text traversal. We adopted a similar approach, as shown in Fig. \ref{fig:gaze_scanpath}, using a blue-to-red gradient to represent the temporal progression of the scanpath. While this provided a useful overview of gaze patterns, it was still challenging to capture a detailed view of gaze movements over time. Therefore, we developed gaze scanplot videos that show gaze movements with corresponding timestamps, enabling replay and a more in-depth examination of the scanpaths.

\section{Methodology}\label{sec:method}

We employed a mixed-method approach to analyze and compare instructed versus in-the-wild reading behaviors by using quantitative and qualitative analysis methods on the gaze scanpaths. The analysis begins with a hand-coding process of the in-the-wild gaze scanpaths. Through our human-driven exploration, we were able to identify naturally-emerging reading behaviors. Reviewers segmented and labeled gaze scanpath videos, refining the taxonomy as new behaviors were identified and resolving disagreements through an inter-rater reliability (IRR) process. This annotated data then informed the training of AI models for automated behavior recognition. A baseline was established with a bag-of-classifiers approach, while 1D and 2D CNNs were developed for higher-resolution predictions. This approach provides a robust framework for understanding and classifying natural reading behaviors in real-world educational contexts.

\subsection{Human Annotation}

Observations of the scanplot images and videos revealed a range of behaviors, including conventional ``zigzag'' reading, skimming, and re-reading. To distinguish and analyze these varied behaviors, we conducted a human-driven annotation process to segment and classify the students' coldread gaze scanpaths. Each session was annotated by two of the three trained human reviewers, with the third reviewer available to resolve any labeling disagreements. The reviewers were trained and knowledgeable in eye-tracking data and visualizations.

\subsubsection{Annotation Protocol}

As identified in our literature gap, a consistent taxonomy for reading behaviors (e.g., ``skimming'', ``deep'', ``scanning'') has not been established. Different studies use varying and often overlapping labels, with definitions tied to specific cognitive tasks. To address these challenges, we initiated our annotation process by adopting \citet{Campbell2001}'s ``regular'' and ``skimming'' labels and definitions as our starting point. The annotation team agreed that, throughout our review of the data, any encounter with unlabeled or unseen behaviors would necessitate creating a new label, saving an instance of the behavior as a reference example, and revisiting previous sessions to identify and update labels for any occurrences of this new behavior. This iterative process of annotating new samples, identifying novel behaviors, and refining our taxonomy and framework aimed to capture and measure in-the-wild reading behaviors accurately.

\begin{figure}[h]
    \centering
    \begin{subfigure}[b]{\textwidth}
        \centering
        \includegraphics[width=0.85\textwidth, trim={0cm 2.5cm 0 3.25cm},clip]{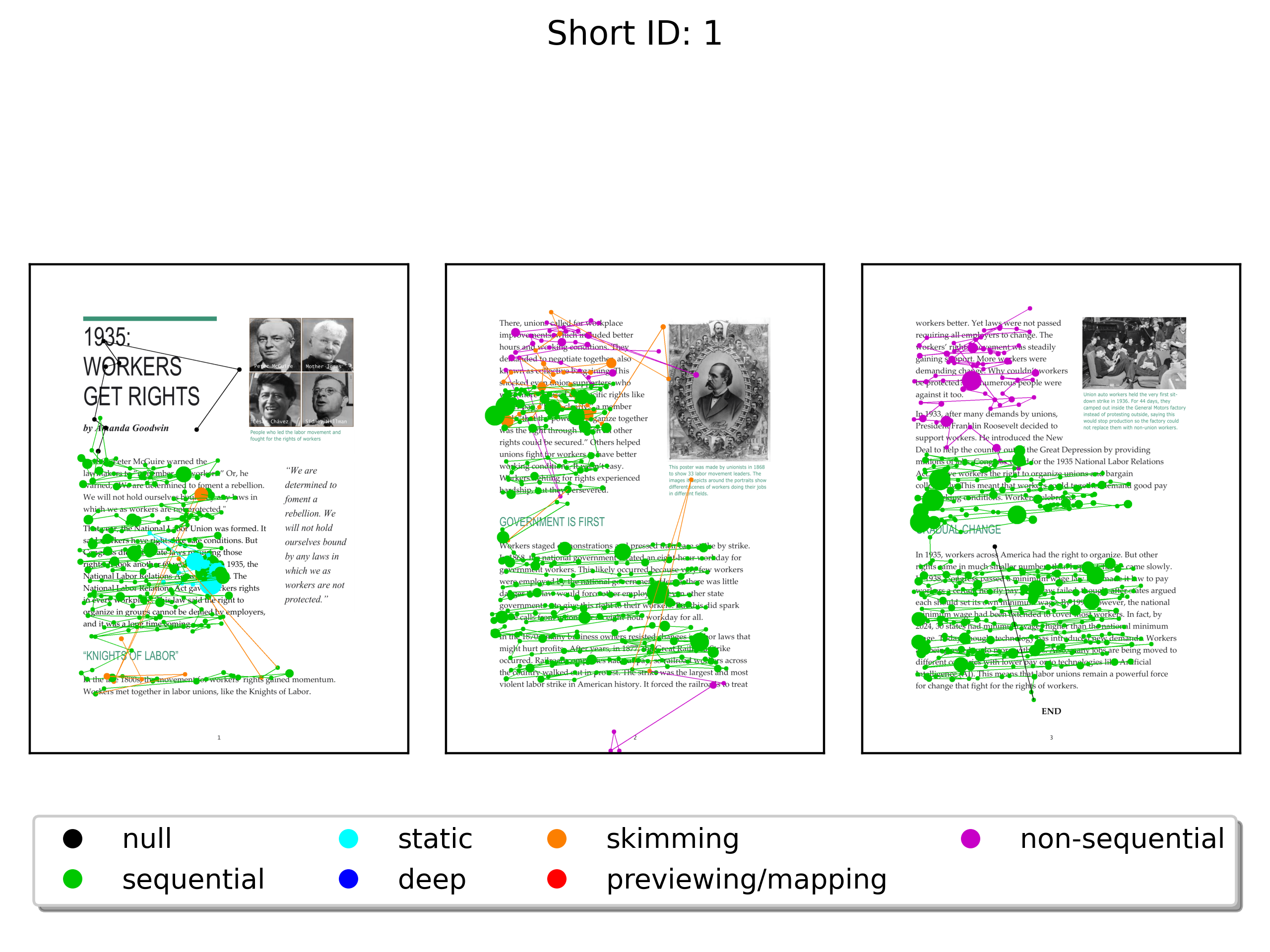}
        \label{fig:example_1}
    \end{subfigure}
    \newline
    \begin{subfigure}[b]{\textwidth}
        \centering
        \includegraphics[width=0.85\textwidth, trim={0 2.5cm 0 3.25cm},clip]{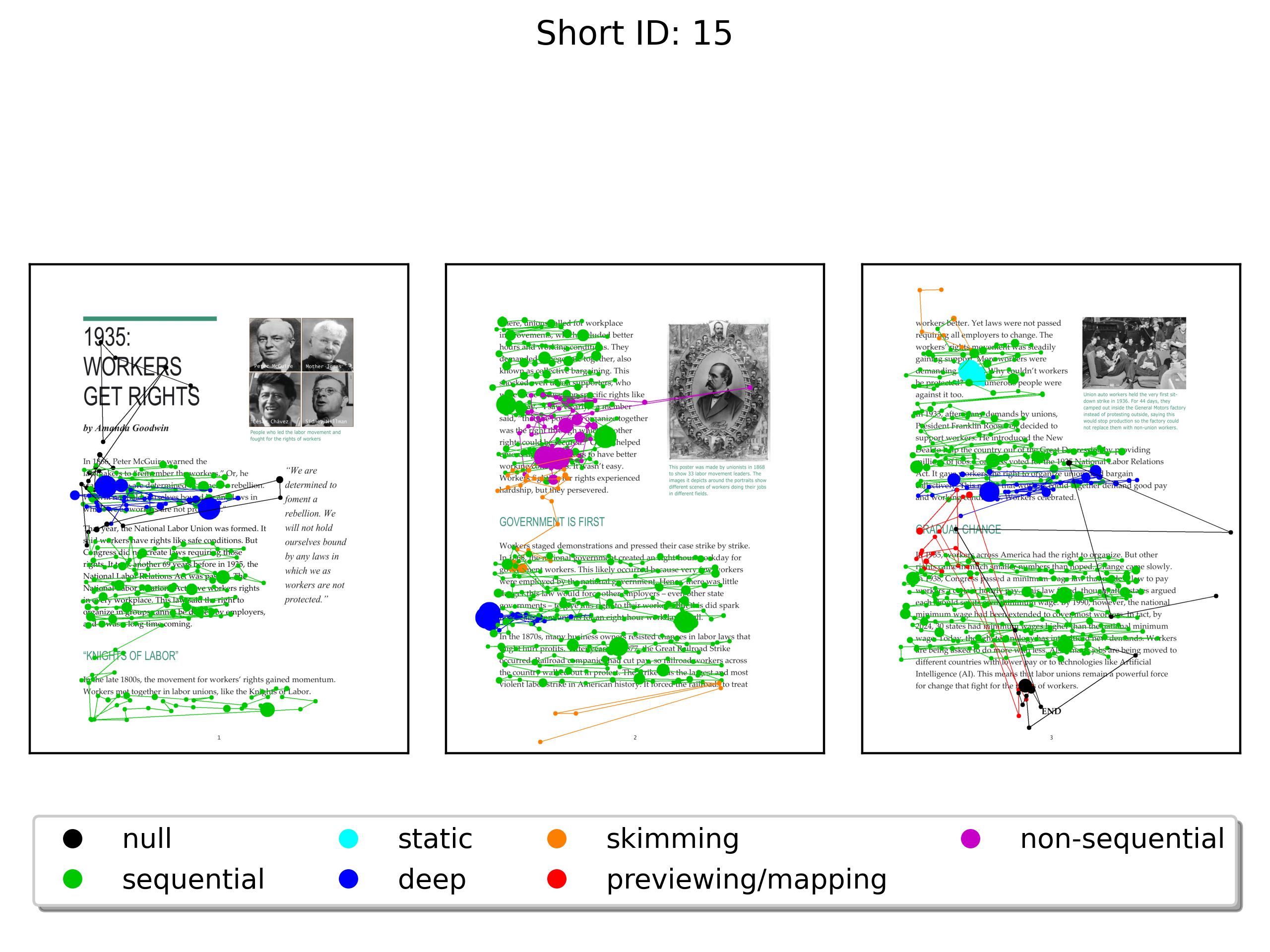}
        \label{fig:example_2}
    \end{subfigure}
    \newline
    \begin{subfigure}[b]{\textwidth}
        \centering
        \includegraphics[width=0.85\textwidth, trim={0cm 0 0 3.25cm},clip]{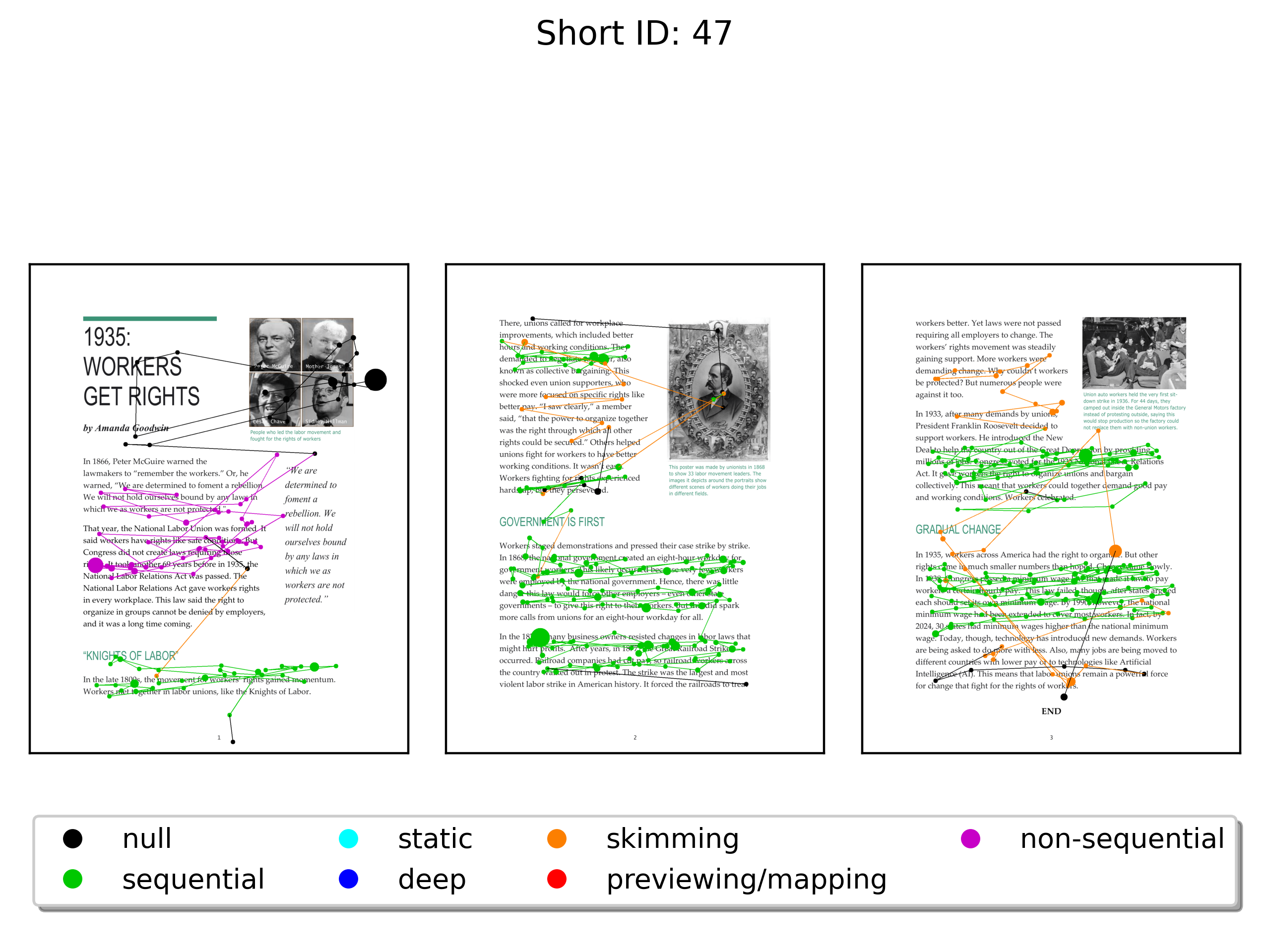}
        \label{fig:example_3}
    \end{subfigure}
    \Description{Three example human-labeled in-the-wild behaviors that have been overlaid on top of the PDF text. In these examples, we see that students use many different reading behaviors to navigate the text.}
    \caption{\textbf{Example Human-Labeled In-the-Wild Behaviors}: Gaze scanpaths from participants 1 (\textbf{top}), 15 (\textbf{center}), and 47 (\textbf{bottom}). illustrating how different reading behaviors are leveraged to explore and comprehend the text.}
    \label{fig:examples}
\end{figure}

In our annotation process, we utilized both the accumulative gaze scanpath figures and replay video to identify and recognize reading behaviors. To avoid different start and end times for behavior segments by different reviewers that could complicate the calculation of inter-rater reliability, it was decided that a single reviewer would first segment the video and assign an initial label. The second reviewer would then examine the predefined segment and independently assign their label without knowledge of the first. If the second reviewer considered that the original segmentation encapsulated more than one behavior, the original segmentation was further spliced and annotated during a collaborative session with both reviewers. An example annotation table is presented in Table \ref{fig:annotation_table}. A final label was determined if both reviewers agreed. Any disagreements were resolved in IRR round sessions, where each reviewer could justify their label, followed by a re-vote to finalize the label. Additional metadata, such as the words covered and the word per minute (WPM), was recorded. This differentiation prevents the overuse of the term WPM which is associated with the reading speed rather than the skimming speed of an individual. Fig. \ref{fig:examples} illustrates three examples of human-labeled in-the-wild reader behaviors.

\begin{table}[ht]
    \begin{tabular}{rrlllrr}
    \hline
    Start Time (mins) & End Time (mins) & Label 1       & Label 2       & Final Label   & \makecell{Words \\ Covered} & WPM \\
    \hline
    0:12                                 & 0:36                               & deep           & deep           & deep           & 10                                 & 25.32                    \\
    1:20                                 & 1:56                               & regular        & regular        & regular        & 60                                 & 99.26                    \\
    1:56                                 & 2:2                                & deep           & deep           & deep           & 2                                  & 21.2                     \\
    2:2                                  & 2:28                               & regular        & regular        & regular        & 34                                 & 80.31                    \\
    2:28                                 & 2:36                               & skimming       & non-sequential & skimming       & 25                                 & 178.57                   \\
    2:36                                 & 2:52                               & non-sequential & non-sequential & non-sequential & 49                                 & 186.43                   \\
    2:52                                 & 3:6                                & regular        & non-sequential & regular        & 30                                 & 124.14                   \\
    3:13                                 & 3:26                               & regular        & regular        & regular        & 19                                 & 94.76                    \\
    \hline
    \end{tabular}
    \Description{This table provides an example of the annotation process used to segment and classify reading behaviors based on gaze scanpath videos. Each row includes the start and end times of a reading segment (in minutes), two initial labels from annotators, a final agreed-upon label, the number of words covered during the segment, and the calculated words per minute (WPM). The behaviors classified include "deep," "regular," "skimming," and "non-sequential," highlighting the variety and complexity of reading patterns observed.}
    \caption{\textbf{Example Annotation Table}: Our annotation process involved the segmentation and classification of the reading behaviors with respect to the gaze scanpath videos. Therefore, a start time and end time along with a label were the key fields to annotate.}
    \label{fig:annotation_table}
\end{table}

\subsubsection{Inter-rater Reliability}

We began with an initial practice round using the ``regular'' and "skimming'' labels from \citet{Campbell2001} to train all reviewers on the annotation process. Behavior archetypes, illustrated in Fig. \ref{fig:example_behavior}, were documented and utilized as references to help reviewers validate and communicate labels consistently. During the second IRR round, we began identifying unexpected behaviors that did not visually align with the ``regular'' and ``skimming'' descriptions. One such behavior was ``non-sequential'' reading, where participants continued through the passage but did not follow the conventional ``zigzag'' pattern. Instead, the gaze scanpath diverged significantly from the passage's order, creating a star-like pattern. This behavior was labeled as ``non-sequential'' reading. To better differentiate ``regular reading'' from ``non-sequential'', we renamed ``regular reading'' to ``sequential''. Additionally, two slow-moving behaviors were observed. The first, labeled as ``deep'', represented a much slower version of ``sequential'' reading, characterized by a ``zigzag'' pattern that suggested careful, word-by-word re-reading within lines. We hypothesized that this behavior occurs when students struggle with understanding a word or sentence and engage in deep reading to enhance comprehension. The second slow-paced behavior, labeled as ``static'', was identified when the gaze remained stationary for more than approximately 5 seconds.

In the third IRR round, we identified only one new behavior, another fast-paced behavior that appeared different from ``skimming''. This behavior, labeled as ``previewing/mapping'', was characterized by swift movements with large saccades across the entire text. It appeared that students used this behavior to self-pace and map out the text's layout. ``previewing/mapping'' was most commonly observed at the beginning of pages or paragraphs, with quick glances toward the bottom of the text. This was distinct from ``skimming'', where the gaze scanpath still traversed the text quickly, with fixations across paragraphs to grasp content. In our final round, no new behaviors were identified. Cohen's Kappa \cite{CohensKappa} is a statistical measure for assessing inter-rater reliability, accounting for chance agreement between annotators. As shown in Table \ref{tab:irr_rounds}, the Cohen's Kappa for our third round—the first large-scale annotation round was 0.57. This was improved to 0.65 by the fourth round. Through our discussions during these sessions, we found that many disagreements arose when annotating transitional or intermediary behaviors. Examples included a rapid ``sequential'' read that shifted into a full skim, or a noisy ``sequential'' read with many interruptions, which could appear similar to ``non-sequential'' or ``previewing/mapping'' behaviors but still exhibited a distinct ``zigzag'' pattern.

To enhance the reliability and validity of our work, we implemented a procedure of double-coding all behaviors, resolving disagreements through discussion. During the annotation process, reviewers collaboratively developed and utilized a shared document containing visual examples and detailed descriptions of reading behaviors. This document aided consensus and a shared understanding of the nuanced behaviors observed, while the iterative IRR process further trained annotators and helped identify additional behaviors, aligning with the study’s focus on capturing the complexity of in-the-wild reading behaviors.

\begin{table}[ht]
    \begin{tabular}{c c c c m{5.5cm} }
        \hline
        Round & \makecell{Number of \\ Sessions} & \makecell{Cohen's \\ Kappa} & Added Labels & Comments \\
        \hline
        1 & 1 & NA & "regular reading'' and "skimming'' & Following \citet{Campbell2001} behavior taxonomy as a baseline.\\
        2 & 2 & 0.73 & ``deep'', ``non-sequential'', ``static'' & "regular reading'' is redefined as ``sequential'' to be more precise.\\
        3 & 11 & 0.57 & "previewing/mapping'' & \\
        4 & 14 & 0.65 & NA & No new unlabeled behaviors \\
        \hline
    \end{tabular}
    \Description{This table summarizes the iterative rounds of inter-rater reliability (IRR) sessions conducted during the annotation process. Each row details the round number, the number of sessions reviewed, Cohen's Kappa score indicating inter-rater agreement, any new labels introduced during the round, and relevant comments. Labels such as "regular reading," "skimming," and "deep" were refined or added in earlier rounds, with additional behaviors like "non-sequential," "static," and "previewing/mapping" introduced as the process progressed. The table highlights how disagreements were resolved collaboratively, and how the taxonomy evolved through successive rounds.}
    \caption{\textbf{IRR Rounds}: Each IRR Round was a session where all 3 reviewers met, discussed disagreements, collectively watched the gaze scanpath video, and voted on the final label. New behavior labels were created if a new behavior was visually too distinct from prior labels and their accompanying example snapshots.}
    \label{tab:irr_rounds}
\end{table}

\subsection{Behavior Classifier}

To demonstrate the feasibility of using AI to recognize complex and fast-paced reading behaviors, we trained a model using both accumulative and per-behavior statistics derived from the in-the-wild corpus. Tab. \ref{tab:per_behavior_stats} provides a detailed summary of per-behavior statistics, including key metrics that characterize the dataset. Tab. \ref{tab:corpus_overview} provides an overview of the corpus characteristics, while Fig. \ref{fig:corpus_figures} illustrates accumulative and per-behavior counts, durations, and fixations. A significant challenge was the class imbalance, with the ``sequential'' class dominating the dataset, which limited the ability to represent other behaviors adequately. Furthermore, the brief duration of many behaviors introduced complexity in the segmentation process. Recognizing these constraints, we focused on the ``sequential'', ``non-sequential'', and ``skimming'' behaviors for training, as the limited number of instances for other behaviors restricted their inclusion. Rather than aiming to develop a generalizable model, this effort serves as a proof-of-concept, illustrating the potential of AI for identifying nuanced reading behaviors.

\begin{table}[h]
    \begin{tabular}{lllll}
    \hline
    Behavior           & Median Duration (s) & \makecell{Median Fixation \\ Count} & \makecell{Median Scanpath \\ Length (cm)} & \makecell{Median Fixation \\ Duration (ms)} \\
    \hline
    static             & 4.98 (2.31)         & 11.00 (3.00)          & 8.24 (4.92)                 & 467.03 (247.04)               \\
    deep               & 11.61 (6.16)        & 25.00 (12.25)         & 49.74 (35.83)               & 262.02 (100.46)               \\
    sequential & 25.84 (29.15)       & 63.00 (80.00)         & 141.03 (187.97)             & 232.55 (51.92)                \\
    non-sequential     & 8.68 (9.86)         & 24.00 (17.50)         & 54.66 (72.44)               & 221.60 (57.67)        \\       
    skimming           & 5.80 (5.30)         & 10.00 (11.00)         & 30.59 (37.02)               & 181.67 (47.02)                \\
    previewing/mapping & 7.50 (6.55)         & 8.00 (12.50)          & 51.28 (53.01)               & 208.70 (55.46)                \\
    \hline
    \end{tabular}
    \Description{This table provides median descriptive statistics for in-the-wild reading behaviors, including duration, fixation count, scanpath length, and fixation duration. Median values are accompanied by interquartile ranges (IQRs) to highlight variability across behaviors such as ``static'', ``deep'', ``sequential'', ``non-sequential'', ``skimming'', and ``previewing/mapping.''}
    \caption{\textbf{Per-Behavior Statistics for In-the-Wild Reading Behaviors}: Descriptive statistics are presented for each behavior. Median values are reported alongside interquartile ranges (IQRs) to account for variability and skewness in the data.}
    \label{tab:per_behavior_stats}
\end{table}

\newcommand{\figureSize}{0.45}
\begin{figure}[h]
    \centering
    \begin{subfigure}[b]{\figureSize\textwidth}
        \centering
        \includegraphics[width=\textwidth]{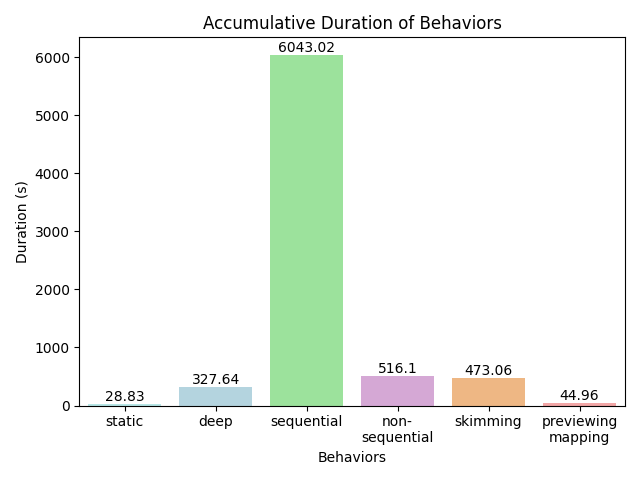}
        \label{fig:corpus_accumulative_behavior_duration}
    \end{subfigure}
    \hfill
    \begin{subfigure}[b]{\figureSize\textwidth}
        \centering
        \includegraphics[width=\textwidth]{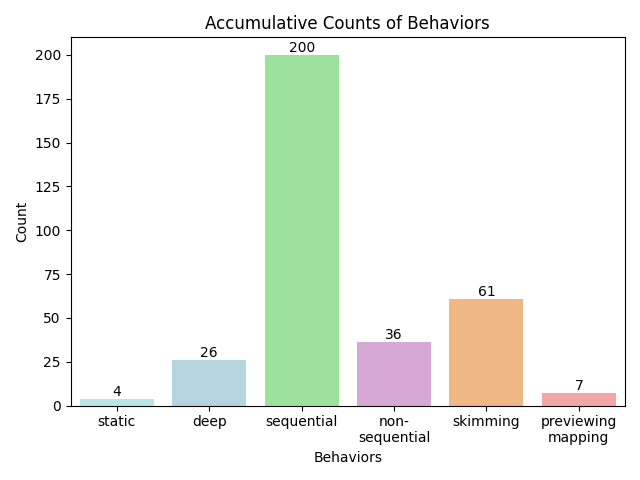}
        \label{fig:corpus_accumulative_counts}
    \end{subfigure}
    \newline
    \begin{subfigure}[b]{\figureSize\textwidth}
        \centering
        \includegraphics[width=\textwidth]{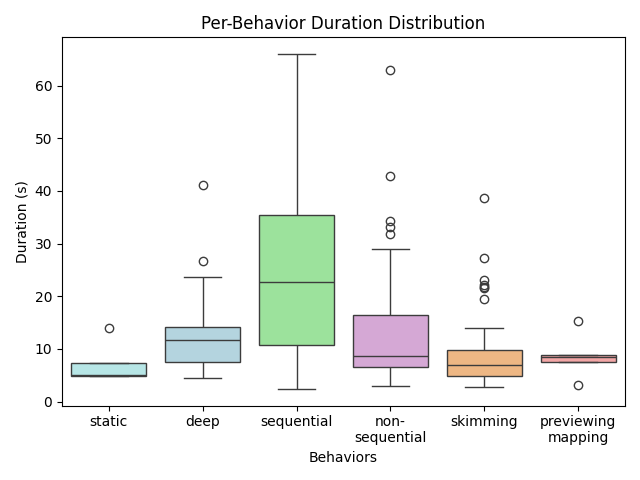}
        \label{fig:per_behavior_corpus_duration}
    \end{subfigure}
    \hfill
    \begin{subfigure}[b]{\figureSize\textwidth}
        \centering
        \includegraphics[width=\textwidth]{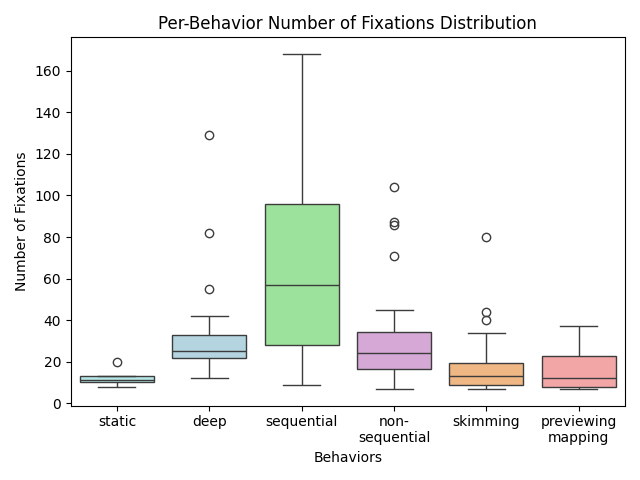}
        \label{fig:per_behavior_corpus_fixation_count}
    \end{subfigure}
    \Description{Four plots describing the statistical attributes of In-the-Wild Dataset. The first two plots contain accumulative metrics, with a large imbalance of fixation counts and durations across ``static'', ``deep'', ``sequential'', ``non-sequential'', ``skimming'', and ``previewing/mapping''. ``sequential'' reading has approximately 10x more samples many other behaviors.}
    \caption{\textbf{In-the-wild Behavior Corpus Characteristics}: Duration, count, and fixation count for accumulative and per-behavior. From these plots, we can observe the significant class imbalance favoring ``sequential'' reading. Lastly, we visualize the duration and fixation length distributions per behavior to further point how many of these in-the-wild reading behaviors are swift.}
    \label{fig:corpus_figures}
\end{figure}

\begin{table}[h]
\begin{tabular}{ll}
\hline
Metric                     & Value    \\
\hline
Total Session Duration (s) & 7,629.44 \\
Total Fixation Count       & 18,713   \\
Total Behaviors            & 334      \\
Avg. Behaviors/Session     & 12.37    \\
Avg. Duration/Session (s)  & 282.57  \\
\hline
\end{tabular}
\caption{\textbf{Dataset Overview}: Summary of key metrics for the in-the-wild reading behavior dataset.}
\Description{This table summarizes key metrics of the in-the-wild reading behavior dataset, including total session duration, total fixation count, total behaviors, average behaviors per session, and average session duration.}
\label{tab:corpus_overview}
\end{table}

Since the corpus consists of scanpath time-series sequences that are not independent and identically distributed (IID) \cite{Hsieh2020}, we opted for a leave-one-participant-out cross-validation (LOPOCV) approach to assess the model's effectiveness and generalizability on unseen data. LOPOCV presents a more rigorous cross-validation method than a random train-validate-test split, as it prevents the model from being exposed to any sequence examples from the test participant during training. For our initial approach, we built upon the existing literature that employs a sliding time-window technique to compute time-accumulative gaze metrics, which are then used as input features for machine learning algorithms \cite{Chen2023, Srivastava2018CombiningRecognition, BektaGEAR_MANUAL, Kiefer2013UsingMaps, ZurichEyeRecognition}. However, a major drawback of this method is that smaller time-window sizes can negatively impact performance. Conversely, increasing the window size could result in multiple labels appearing within a single window, creating ambiguity over which label should serve as the ground truth. One potential solution is to select the dominant class within the window, but this would further skew the dataset and could misleadingly suggest that the model is performing better than it actually is. Enlarging the window size would shift the dataset disproportionately toward the majority class, ``sequential'', potentially resulting in a falsely ``perfect'' model. Thus, we constrained the time-window size to a range of $[2, 15]$ seconds based on the per-behavior duration distribution, as shown in Fig. \ref{fig:corpus_figures}, to ensure that the window size does not exceed the mean duration of most behaviors.

\subsubsection{Bag-of-Classifiers}
We employed a bag-of-classifiers from the Python scikit-learn package \cite{scikit-learn} for training ML algorithms with a LOPOCV approach. Within each time window, we computed the following accumulative gaze metrics: (1) fixation count, (2) mean fixation duration, (3) fixation dispersion, (4) mean saccade length, (5) saccade vertical next target, (6) saccade horizontal later, (7) saccade line regression rate, and (8) saccade regression rate. We follow the mathematical definitions introduced in \citet{Busjahn2015EyeOrder} to compute our input features to the ML algorithms. We sweep across $t=[2, 15]$ time-window sizes. Overall, the performance of ML classifiers increases with large time-window sizes.


\subsubsection{CNN Implementation Details} 


Rather than using an accumulative feature time-window approach, more recent methods in time-series analysis leverage CNNs \cite{LongRethinkingBaseline} and long short-term memory (LSTM) networks \cite{Elsworth2020}. Given that gaze scanpath data is not time-independent and that both past and current behaviors can influence future predictions, an LSTM approach could be ideal for capturing these temporal dependencies. However, we opted for a CNN-based approach with a short fixation sequence window. The decision to use a CNN was based on its data efficiency; training an LSTM typically requires a larger corpus, can lead to error accumulation over time, and is prone to significant bias errors when working with smaller datasets \cite{Sun2024OnModels}.

\begin{figure}[h]
    \centering
    \begin{subfigure}[b]{0.329\textwidth}
        \centering
        \includegraphics[width=\textwidth]{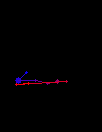}
        \label{fig:cnn_example_regular}
    \end{subfigure}
    \hfill
    \begin{subfigure}[b]{0.329\textwidth}
        \centering
        \includegraphics[width=\textwidth]{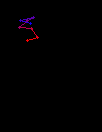}
        \label{fig:cnn_example_non_sequential}
    \end{subfigure}
    \hfill
    \begin{subfigure}[b]{0.329\textwidth}
        \centering
        \includegraphics[width=\textwidth]{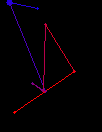}
        \label{fig:cnn_example_skimming}
    \end{subfigure}
    \Description{This figure shows three examples of 2D gaze scanpath input images used for the CNN model, displayed on a black background. The images represent different reading behaviors: sequential reading on the left, non-sequential reading in the center, and skimming on the right. The scanpaths use a blue-to-red gradient to indicate the progression of time, and each image is rendered on a standard letter-size frame to standardize input dimensions.}
    \caption{\textbf{Model Input Examples of 2D Gaze Scanpaths:} ``sequential'' reading on the \textbf{left}, ``non-sequential'' reading on the \textbf{center}, and "skim'' reading on the \textbf{right}. These scanpath input images are generated from a fixation window that is rendered on a standard letter-size image. Similar to the gaze scanpaths that human reviewers used to annotate behaviors, a blue-to-red gradient is used to represent the time dimension.}
    \label{fig:cnn_instructed_behaviors}
\end{figure}

Inspired by \citet{Cole2021ConvolutionalMovements}, we developed two CNN-based approaches: a 1D CNN and a 2D CNN. The 1D CNN model takes as input the raw $xy$ fixation subsequence, while the lightweight 2D CNN employs a pretrained ResNet18 backbone \cite{He2016} with three fully connected layers. This 2D model processes scanplot images generated from scanpath subsequences, examples are shown in Fig. \ref{fig:cnn_instructed_behaviors}. Both methods utilize fixation windows of length 10, allowing for much finer behavior predictions. To compute the classification loss, a standard cross-entropy loss function was used. The CNN models were trained using the Adam optimizer for 50 epochs, with an early stopping mechanism to prevent overfitting. Gaussian noise was add to the input fixation xy points as a data augmentation technique to improve model performance. We performed an exhaustive grid search to identify the ideal hyperparameters for both 1D and 2D CNN including: learning rate, weight decay rate, model design, batch size, and scanpath plot generation.

\section{Results}\label{sec:results}

We present gaze scanpath examples of both instructed and in-the-wild behaviors to illustrate their distinctions. As our gaze-metric data does not follow a normal distribution and remains robust to variability, we use Mann-Whitney $U$ tests \cite{MannUWhitney}, a non-parametric statistical method used to compare differences between two independent groups without assuming a normal distribution. To account for multiple comparisons, we apply Bonferroni corrections to compare instructed and in-the-wild conditions. To analyze the differences between ``sequential'' and ``non-sequential'' reading, we use the Independent Two-Sample $t$-test \cite{IndependentTwoTest}, a parametric statistical method that compares the means of two independent groups to determine whether the differences are statistically significant. This test is appropriate here as it assumes that the data within each group is approximately normally distributed and helps quantify the distinct characteristics of these two reading behaviors. Additionally, we employ pairwise Hotelling's $T$-squared tests \cite{Hotelling}, a multivariate statistical method used to compare the means of two groups across multiple variables simultaneously. This test is particularly suited for examining differences between in-the-wild behavior pairs, as it accounts for the interdependence among gaze-metric variables such as velocity, density, and sequentiality, providing a comprehensive assessment of their distinctions. Finally, we report the performance metrics of the 2D CNN model.

\subsection{Contrast between Instructed and In-the-Wild Behaviors}

\begin{figure}[h]
    \centering
    \begin{subfigure}[b]{0.329\textwidth}
        \centering
        \includegraphics[width=\textwidth, trim={4.25cm 1cm 4.25cm 1cm},clip]{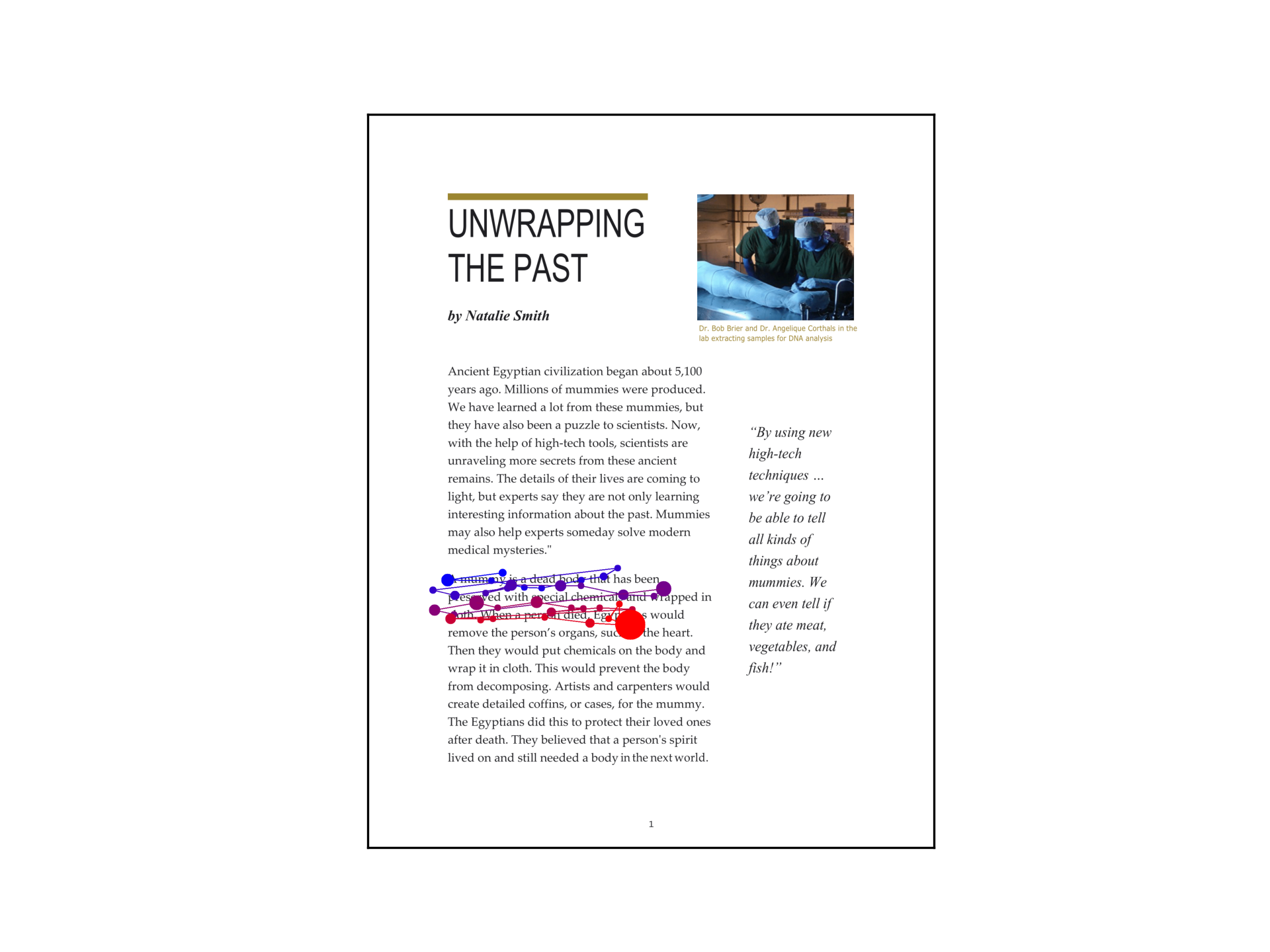}
        \label{fig:example_deep}
    \end{subfigure}
    \hfill
    \begin{subfigure}[b]{0.329\textwidth}
        \centering
        \includegraphics[width=\textwidth, trim={4.25cm 1cm 4.25cm 1cm},clip]{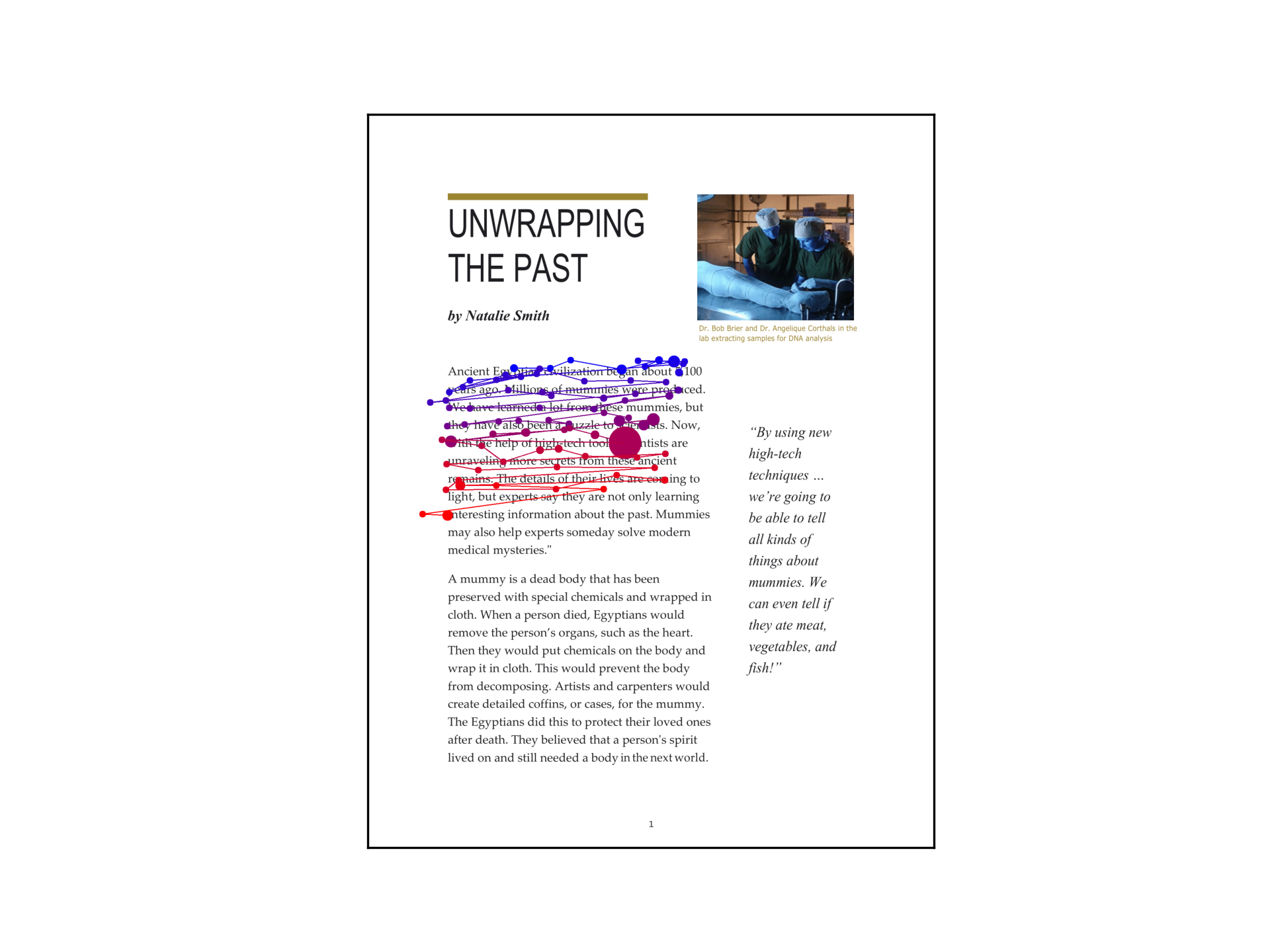}
        \label{fig:example_linear}
    \end{subfigure}
    \hfill
    \begin{subfigure}[b]{0.329\textwidth}
        \centering
        \includegraphics[width=\textwidth, trim={4.25cm 1cm 4.25cm 1cm},clip]{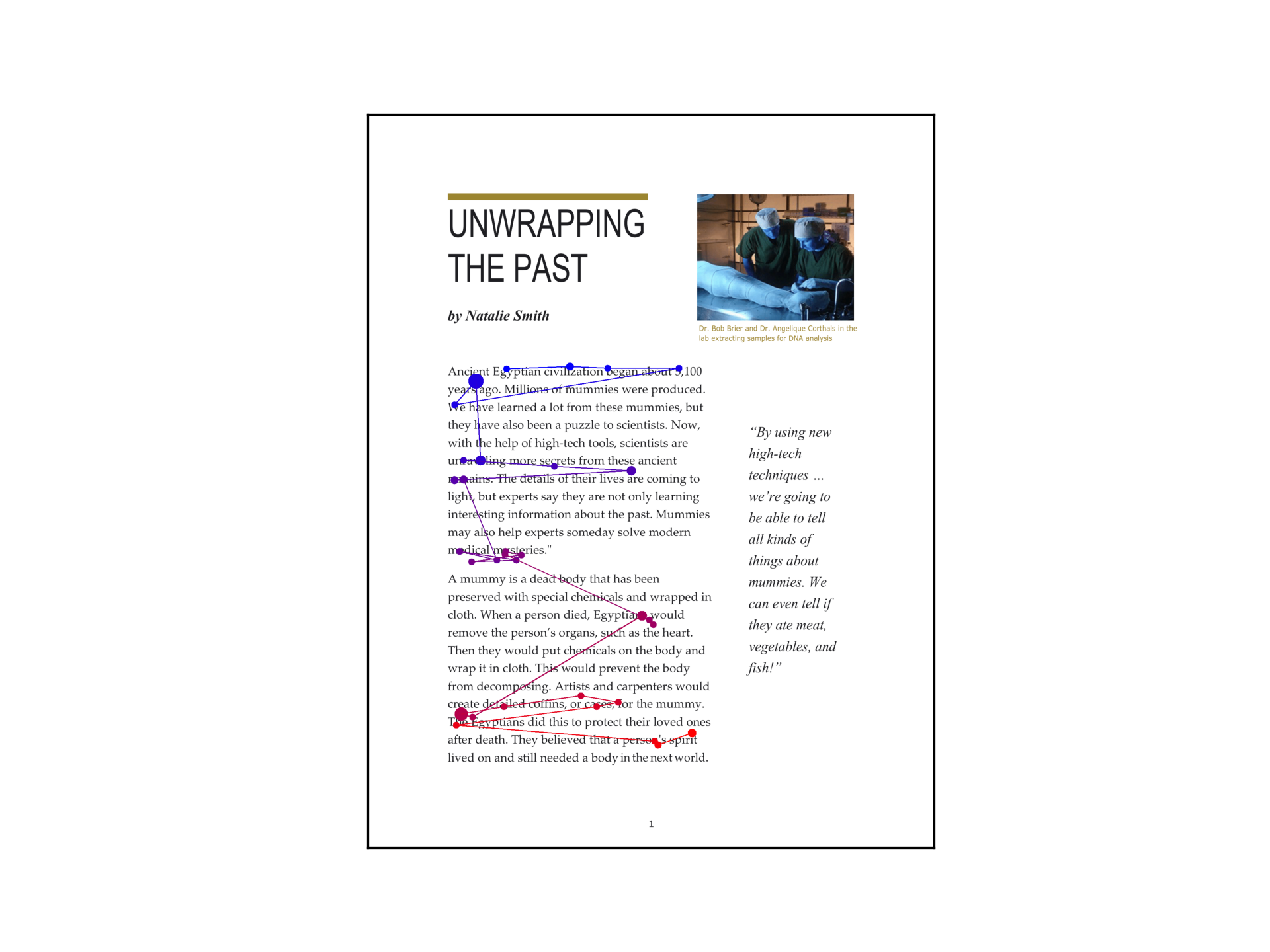}
        \label{fig:example_skimming}
    \end{subfigure}
    \Description{This figure illustrates three examples of gaze scanpaths for instructed reading behaviors. The left image represents "deep" reading, characterized by repeated fixations on a small area of text. The center image shows "regular" reading, with a sequential, line-by-line pattern. The right image depicts "skimming," featuring rapid, broad movements across the text. Each scanpath is visualized with a consistent format, highlighting the distinct eye movement patterns associated with each behavior.}
    \caption{\textbf{Examples of Gaze Scanpaths for Instructed Behavior:} ``deep'' reading on the \textbf{left}, ``regular'' reading on the \textbf{center}, and ``skim'' reading on the \textbf{right}.}
    \label{fig:instructed_behaviors}
\end{figure}

\begin{figure}[h]
    \centering
    \includegraphics[width=\linewidth]{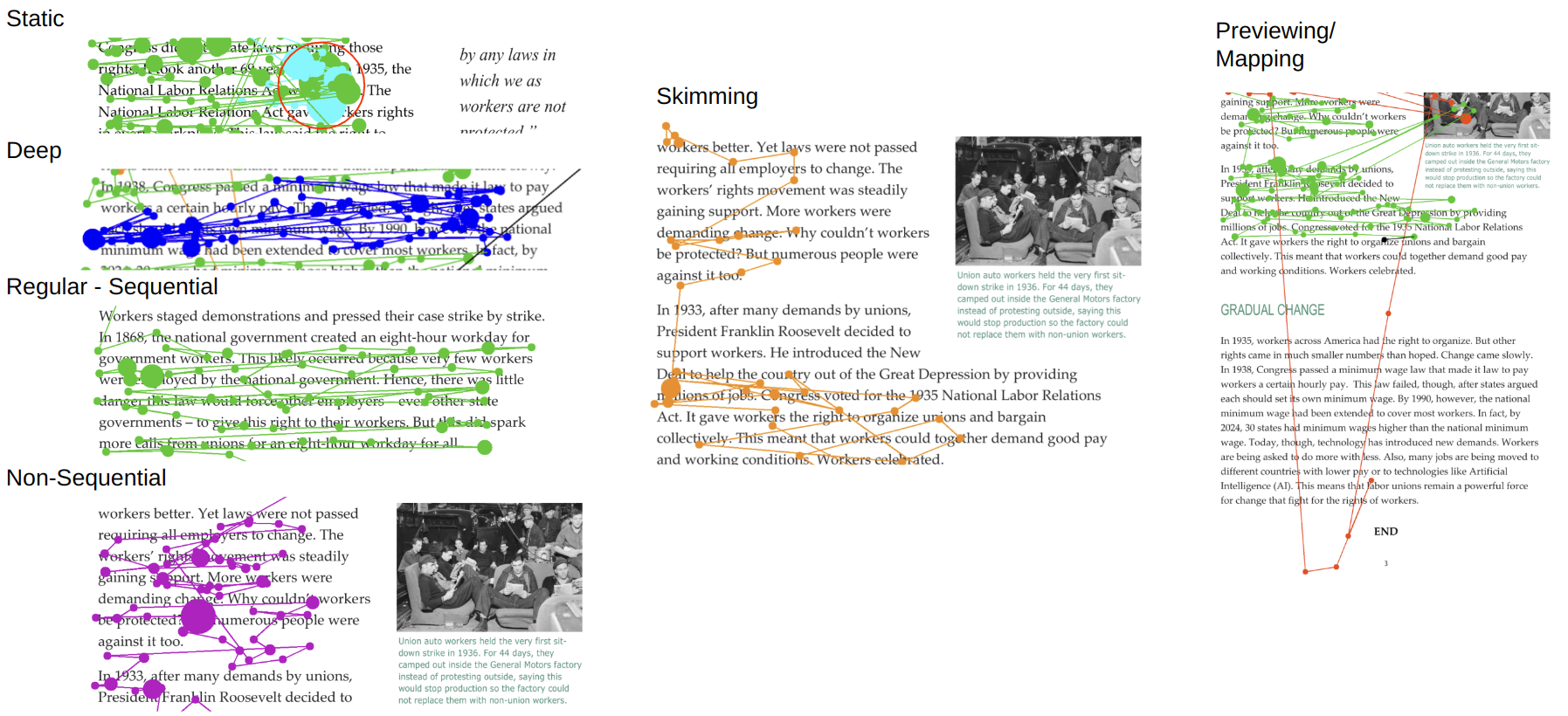}
    \Description{Example scanpath plots of each in-the-wild behavior, with ``static'' having a blob of fixations, ``deep'' having a highly dense ``zigzag'' pattern, ``sequential'' containing a clean line-by-line ``ziggap'', ``skimming'' having fast transitions between parts of the text, and ``previewing/mapping'' having swift movements across the entire document.}
    \caption{\textbf{In-the-Wild Behavior Archetypes:} During our human-driven annotation process, recorded examples of the behaviors as a reference to have a visual representation to compare against new behaviors. Example scanpath plots of each in-the-wild behavior, with ``static'' having a blob of fixations, ``deep'' having a highly dense ``zigzag'' pattern, ``sequential'' containing a clean line-by-line ``ziggap'', ``skimming'' having fast transitions between parts of the text, and ``previewing/mapping'' having swift movements across the entire document.}
    \label{fig:example_behavior}
\end{figure}

Our observations of the instructed behaviors showed that students performed the requested tasks, resulting in clean gaze scanpaths, as illustrated in Fig. \ref{fig:instructed_behaviors}. However, the ``perfection'' of this gaze data limits our ability to capture the natural noise and variability that our theoretical, statistical, and AI models need to account for. Furthermore, instructing students on how to behave based on our hypothesis, rather than observing authentic reading behaviors, restricted our ability to understand naturally occurring reading behaviors. In contrast, the human-driven annotation process of the in-the-wild task allowed us to identify behavioral archetypes, including newly observed behaviors like ``non-sequential'', ``static'', and ``previewing/mapping''. Examples of these in-the-wild behaviors are shown in Fig. \ref{fig:example_behavior}.

\begin{figure}[!ht]
    \centering
    \begin{subfigure}[b]{0.49\textwidth}
        \centering
        \includegraphics[width=\textwidth]{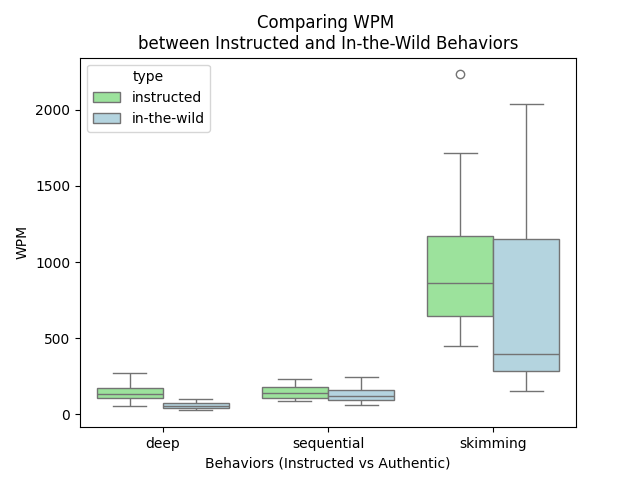}
        \label{fig:WPM_behaviors}
    \end{subfigure}
    \hfill
    \begin{subfigure}[b]{0.49\textwidth}
        \centering
        \includegraphics[width=\textwidth]{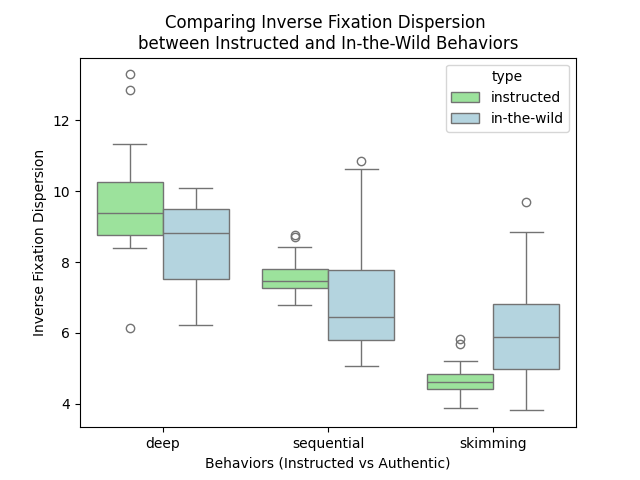}
        \label{fig:density_behaviors}
    \end{subfigure}
    \Description{Two box plot diagrams comparing different reading behaviors across instructed and in-the-wild reading conditions.}
    \caption{\textbf{Gaze Measure Differences Between Instructed and In-the-Wild Behaviors}: The velocity (\textbf{left}) and density (\textbf{right} distributions illustrate how instructed behaviors do not match in-the-wild behaviors.}
    \label{fig:behaviors_differences}
\end{figure}

The box plots in Fig. \ref{fig:behaviors_differences} show WPM and inverse fixation dispersion differences between instructed and in-the-wild reading behaviors. A Mann-Whitney $U$ test showed significant differences for ``sequential'' $(U = 3358, p < 0.001)$ and ``deep'' $(U=326, p < 0.001)$ reading behaviors between instructed and in-the-wild conditions. In contrast, the ``skimming'' $(U = 727, p < 0.348)$ behavior behavior did not show significant differences between conditions.

\subsection{Theoretical Framework and Taxonomy}

\begin{figure}[h]
    \centering
    \begin{subfigure}{0.4\linewidth}
        \centering
        \includegraphics[width=\linewidth]{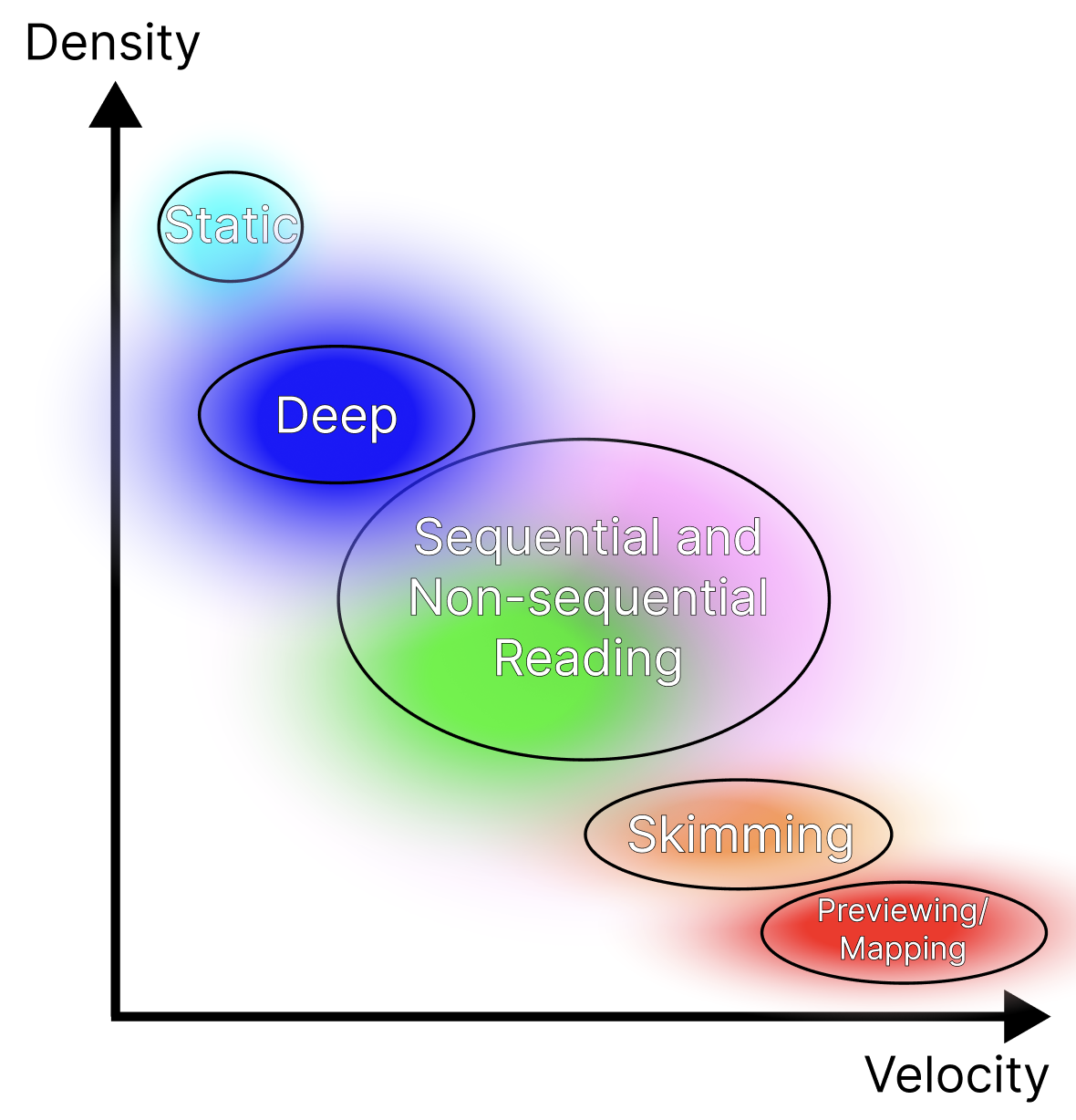}
        \caption{\textbf{Scanpath's density \& velocity as a means to contrast behaviors}: Behavioral regions are placed within the xy plane to characterize and separate behaviors into distinct profiles.}
        \label{fig:framework_diagram_part1}
    \end{subfigure}
    \hfill%
    \begin{subfigure}{0.59\linewidth}
        \centering
        \includegraphics[width=\linewidth]{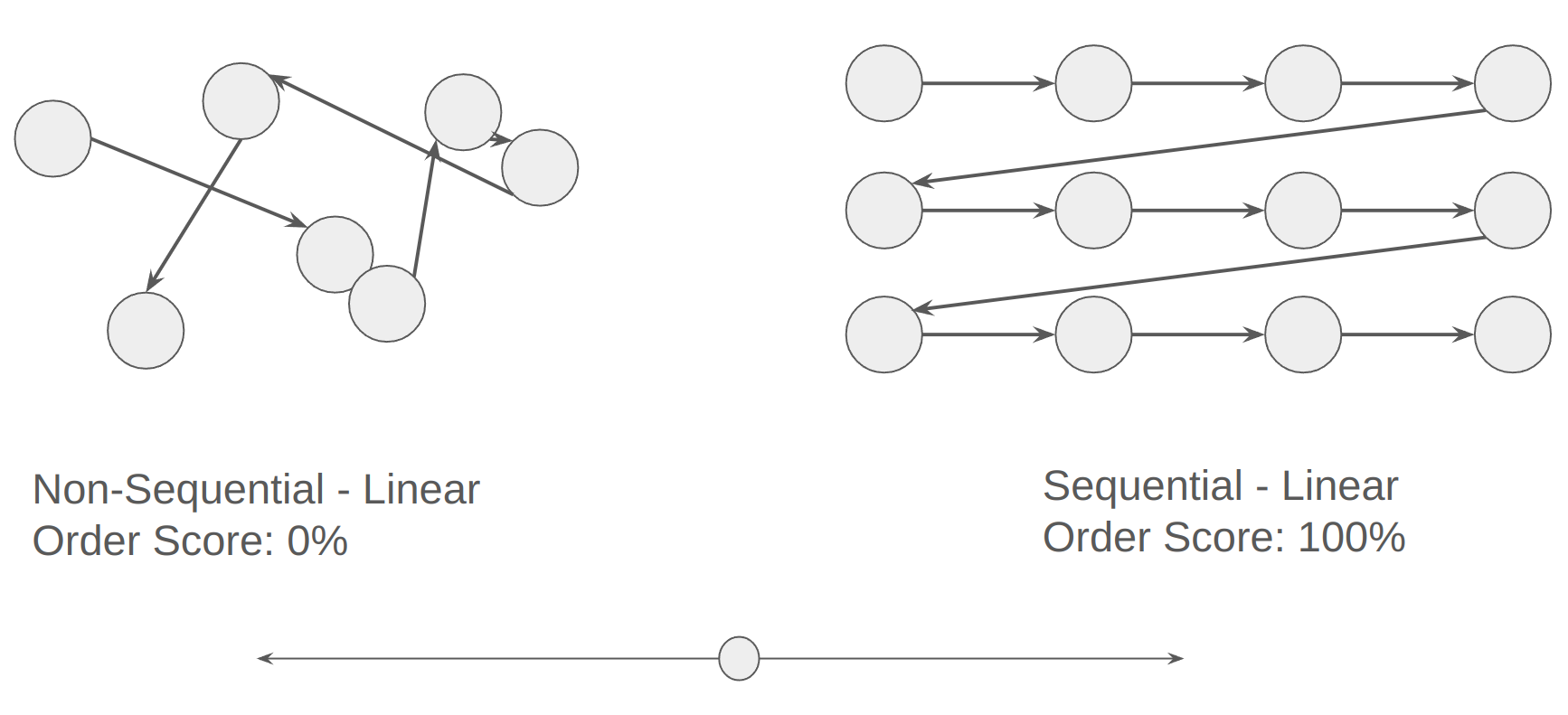}
        \caption{\textbf{Scanpath's text linear order as a sequentiality spectrum:} A sequentiality spectrum with each end being contrasting cases. Zero sequentiality, meaning ``non-sequential'', is random scanpath progressions while 100\% sequentiality follows the text's linear order like a ``zigzag'' pattern.}
        \label{fig:framework_diagram_part_2}
    \end{subfigure}
    \Description{Two spectrum diagrams of physical attributes of gaze. The first diagram is an xy plot with density and velocity as the two axes. Behavioral regions are placed within the xy plot. The second diagram is a line spectrum plot with sequentiality as the axis and each axis ends including archetype examples. Zero sequentiality, meaning non-sequential, is random scanpath directions while 100\% sequential has a clean ``zigzag'' pattern.}
    \caption{\textbf{In-the-Wild Behavior Framework}: Combining density, velocity, and sequentiality of scanpaths as a means to describe gaze-based behaviors. With the combination of these 3 continua, behavior regions, and attributes are defined.}
    \label{fig:framework_diagram}
\end{figure}

Throughout our human-driven annotation process, the language we used to resolve disagreements and describe reading behaviors was instrumental in developing the framework presented in Fig. \ref{fig:framework_diagram}. Reviewers relied on key variables, such as velocity, density, and sequentiality of the gaze scanpaths, when assigning behavior labels and reaching consensus in cases of disagreement. ``static'' behaviors were characterized by stationary gaze scanpaths (near-zero velocity, high density). ``sequential'' and ``non-sequential'' behaviors shared similarities in text progression velocity and density; however, we differentiated them using the sequentiality continuum, which extends the gaze-based measure of linear order proposed by \citet{Busjahn2015EyeOrder}. Behaviors such as ``skimming'' and ``previewing/mapping'' were identified by their rapid and extremely quick bidirectional fixation sequences, with a broad distribution across the text, respectively.

Figure \ref{fig:example_behavior} presents the archetype examples used by annotators to help communication and achieve consensus during the labeling process, with refined descriptions for each behavior detailed below.

\begin{itemize}
    \item \textbf{Static}: This behavior occurs when WPM  approaches zero, and the scanpath remains confined to a small cluster of words for an extended period (approximately 5–10 seconds). While such patterns have been associated with mind wandering in the literature, our study refrains from making such conclusions based on scanpaths alone.
    \item \textbf{Deep}: Deep reading is characterized by the careful re-reading of specific lines or words. Unlike static reading, WPM is notably greater than zero. A key indicator is the repeated traversal of the same line three or more times, signifying a deliberate attempt to comprehend the content in detail.
    \item \textbf{Sequential}: Sequential reading reflects a traditional, line-by-line reading style, often resembling a (potentially noisy) zig-zag pattern. Minor lookbacks are typical and do not disrupt the overall sequential flow. This behavior aligns with standard WPM ranges reported in the literature, accounting for variations in students' age and reading ability. Its primary objective is thorough comprehension of the text.
    \item \textbf{Non-sequential}: Non-sequential reading deviates from the line-by-line pattern. While it shares a similar speed to skimming, its defining feature is its nonlinear trajectory. Unlike skimming, which involves largely forward or backward monotonic movements, non-sequential reading revisits earlier sections of the text within the same paragraph. 
    \item \textbf{Skimming}: Skimming involves rapid and large gaze movements across the text, breaking away from the line-by-line pattern. This behavior is characterized by a fast progression through different parts of the text, aiming to provide a quick overview of its content.
    \item \textbf{Previewing/Mapping}: This behavior is marked by rapid and sharp gaze movements that span the entire page, as students quickly preview and map the document’s structure and layout. These movements are brief and typically focused on understanding the organization of the material rather than its detailed content.
\end{itemize}

\subsection{Velocity, Density, and Sequentiality Continua} \label{sec:continua3D}

\begin{figure}[h]
    \begin{subfigure}{0.49\linewidth}
        \centering
        \includegraphics[width=\linewidth]{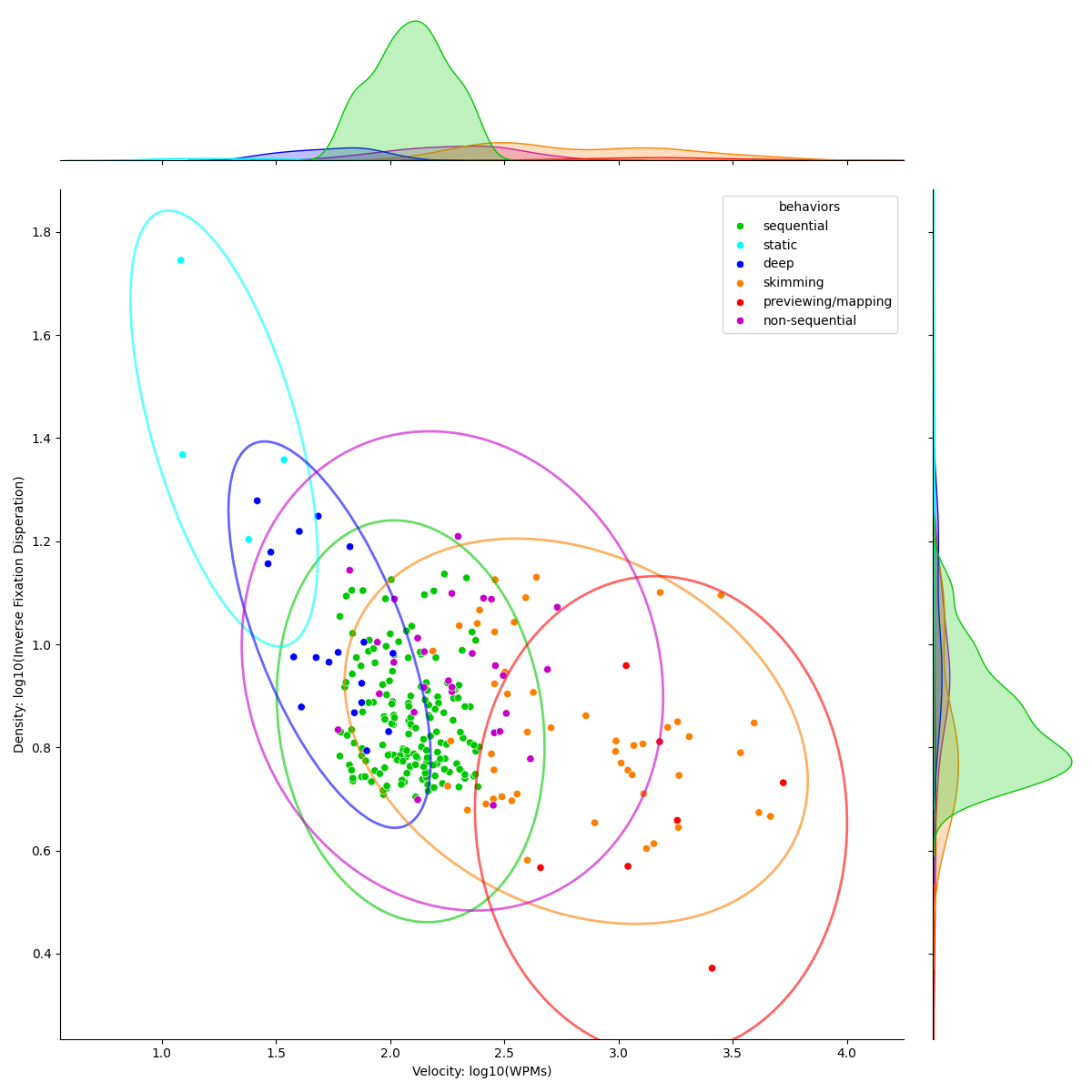}
        \caption{\textbf{Behavior Metrics Density vs. Velocity Plot}: Plotting the per-segment metrics in a logarithmic scale for all behavior types.}
        \label{fig:WPM_vs_density_plot}
    \end{subfigure}
    \hfill%
    \begin{subfigure}{0.49\linewidth}
        \centering
        \includegraphics[width=\linewidth]{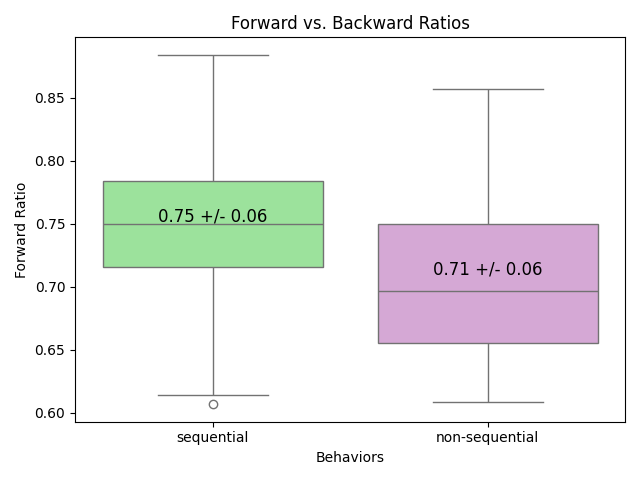}
        \caption{\textbf{Behavior Forward vs. Backward Saccade Ratio for Saccades}: Plotting the per-segment metrics for ``sequential'' and ``non-sequential'' behaviors with their respective mean and standard deviation.}
        \label{fig:fw_bw_ratio}
    \end{subfigure}
    \Description{Two plots illustrate the differences in physical gaze measures across the in-the-wild reading behaviors. The first plot is a xy joint plot in log10-log10 scale with WPM and inverse fixation dispersion as axes. In this plot, we see similar behavior regions as compared to our theoretical framework diagram. In the second plot, box plots comparing ``non-sequential'' and ``sequential'' reading across the forward vs. backward saccade ratio show a difference in mean but with the same variance.} 
    \caption{\textbf{Understanding Behavioral Differences Through Distributions}: These figures use distributions along different spectra to illustrate the differences in reading behaviors.}
    \label{fig:distribution_plots}
\end{figure}

The jointplot in Fig. \ref{fig:WPM_vs_density_plot} shows the WPM and inverse fixation dispersion for in-the-wild reading behaviors in the log10-log10 scale. Pairwise Hotelling's $T$-squared tests \cite{Kariya1981ARobustness} revealed all pairwise comparisons are statistically significant, as shown in Fig. \ref{tab:ptest}, with the exception of ``previewing/mapping'' - ``skimming'' pair with a value of $p=0.194$. This is likely attributed to their scanpath similarity and the small sample size of ``previewing/mapping'' behaviors. The similarity of the jointplot to the theoretical framework and the results of the significance testing further support our behavior regions shown in Fig. \ref{fig:framework_diagram_part1}.

To analyze sequentiality, the box plot in Fig. \ref{fig:fw_bw_ratio} shows the forward vs. backward saccade ratio (FBSR), as described by \citet{Busjahn2015EyeOrder}, of ``sequential'' and ``non-sequential'' in-the-wild behaviors. The ``sequential'' and ``non-sequential'' behavior regions overlap with one another, even though visibly having slightly different distributions in Fig. \ref{fig:framework_diagram_part1}, leverages from the sequentiality continuum to better disambiguate behavior differences. We performed an Independent Two-Sample $t$-test using Python's SciPy package \cite{2020SciPy-NMeth} and identified statistical significance of $p<0.001$; therefore, we can reject the null hypothesis and conclude that the ``sequential'' and ``non-sequential'' behaviors are different along the sequentiality continuum, agreeing with our theoretical framework shown in Fig. \ref{fig:framework_diagram_part_2}.

\begin{table}[h]
    \begin{tabular}{l | llllll}
    \hline
                       & sequential & static   & deep     & skimming & previewing/mapping & non-sequential \\
    \hline
    sequential         & --         & <0.001   & <0.001   & <0.001   & <0.001             & <0.001         \\
    static             &            & --       & 0.0101   & <0.001   & <0.001             & <0.001         \\
    deep               &            &          & --       & <0.001   & <0.001             & <0.001         \\
    skimming           &            &          &          & --       & 0.194              & <0.001         \\
    previewing/mapping &            &          &          &          & --                 & <0.001         \\
    non-sequential     &            &          &          &          &                    & --             \\
    \hline
    \end{tabular}
    \caption{\textbf{Comparing In-the-Wild Behavior via Significance Testing}: Pairwise Hotelling's $T$-squared tests were conducted to evaluate significant differences between the multivariate distributions of behaviors. Adjusted $p$-values are presented in the accompanying matrix, where $p$-values below \textbf{0.05} indicate statistically significant differences between the distributions of the corresponding behaviors.}
    \Description{This table presents the results of pairwise Hotelling's $T$-squared tests for evaluating the statistical significance of differences between multivariate distributions of six reading behaviors: sequential, static, deep, skimming, previewing/mapping, and non-sequential. Adjusted $p$-values are displayed, with values below 0.05 indicating significant differences. Significant pairwise differences are observed for most behavior comparisons, highlighting distinct multivariate distributions across behaviors.}
    \label{tab:ptest}
\end{table}


\subsection{Classification Performance}

\begin{figure}[h]
    \centering
    \begin{subfigure}[b]{0.45\textwidth} 
        \centering
        \begin{tabular}{lllll}
            \hline
            Model               & Recall & Precision & Macro F1 & Accuracy \\
            \hline
            Random              & 0.38   & 0.38      & 0.38     & 0.79     \\
            Majority Class      & 0.59   & 0.53      & 0.56     & 0.87     \\
            SVC                 & 0.59   & 0.53      & 0.56     & 0.87     \\
            1D CNN              & 0.43   & 0.45      & 0.43     & 0.81     \\
            2D CNN              & \textbf{0.81}   & \textbf{0.82}      & \textbf{0.80}     & \textbf{0.92}     \\
            \hline
        \end{tabular}
        \Description{The table compares performance metrics, including Recall, Precision, Macro F1, and Accuracy, for various models applied to the In-the-Wild Behaviors corpus. Models evaluated include Random, Majority Class, SVC, 1D CNN, and 2D CNN. The 2D CNN achieves the highest performance, with a Macro F1 score of 0.80 and an accuracy of 0.92, significantly outperforming other approaches.}
        \caption{\textbf{Comparison with statistical, ML, and CNN models on the In-the-Wild Behaviors corpus:} The ML values reported using a time-window of $t=15$.}
        \label{fig:performance_table}
    \end{subfigure}
    \hfill
    \begin{subfigure}[b]{0.45\textwidth} 
        \centering
        \includegraphics[width=\linewidth]{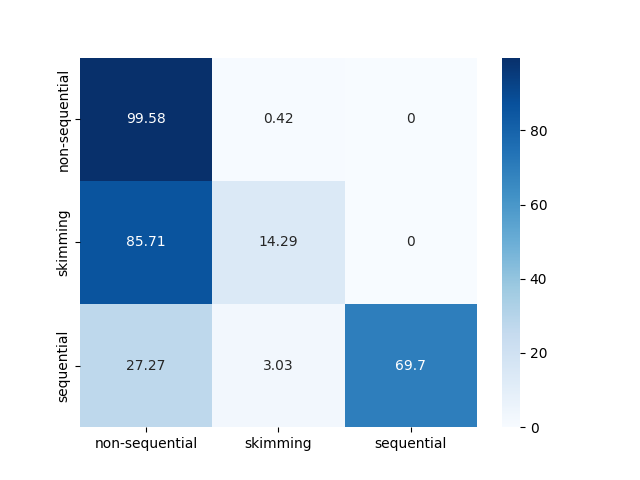}
        \Description{Confusion matrix with strong prediction performance in ``sequential'' and ``skimming'', yet the model misclassified ``sequential'' and ``non-sequential''.}
        \caption{\textbf{Confusion Matrix of the Model Predictions:} Reported the LOPOCV confusion values on the test participants across the In-the-Wild Behavior Dataset. The CNN model has strong prediction performance in ``sequential'' and ``skimming''. The model misclassified ``sequential'' and ``non-sequential''.}
        \label{fig:confusion_matrix}
    \end{subfigure}
    \caption{\textbf{Performance Evaluation of Behavior Classification:} Reporting the performance metrics of ML and CNN-based methods on the \textbf{left} table. The 2D CNN achieved the highest Macro F1 score of 0.8. The confusion matrix for the 2D CNN is shown on the \textbf{right}.}
    \label{fig:combined_figure}
\end{figure}

The results presented in Table \ref{fig:performance_table} highlight the performance of the 2D CNN model compared to traditional machine learning methods such as a Support Vector Classifier (SVC), a majority classifier, and a random classifier. Moreover, the confusion matrix of the 2D CNN models is shown in Fig. \ref{fig:confusion_matrix}. The 2D CNN model achieved a Macro F1 score of 0.80, outperforming all other baseline methods. This demonstrates that the 2D CNN model can more accurately detect and classify in-the-wild reading behaviors by leveraging the temporal dependencies within fixation sequences. Unlike traditional ML methods, which rely on aggregated time-window approaches, the 2D CNN takes a shorter fixation window as input, enabling more nuanced detection of complex and diverse reading behaviors. This reinforces the feasibility of using advanced deep learning methods for real-time, fine-grained analysis of gaze data in dynamic, naturalistic settings.

The statistical models used in Section \ref{sec:continua3D} outperform the AI behavior classifiers primarily due to the nature of the input data and the granularity of analysis. Statistical methods leverage accumulative metrics calculated over the entirety of a behavior segment, such as density and velocity, which provide a holistic, smoothed view of reading behaviors. These metrics aggregate patterns over time, minimizing the influence of transient variations in eye-tracking data. This aggregation process allows statistical models to capture overarching trends and differences between behaviors with greater clarity, resulting in more reliable discrimination across behavior categories.

In contrast, the AI behavior classifiers rely on short time-series windows, which present a much higher level of variability due to transient pauses, uneven gaze movements, and transitions within a single behavior. This variability makes it inherently more challenging for the AI model to learn and generalize behavior patterns, especially when the dataset is small and imbalanced. Furthermore, the limited temporal context within these short windows constrains the model’s ability to identify long-term dependencies or sequential patterns that are readily captured by statistical methods. While the AI classifier demonstrates the feasibility of detecting behaviors in real-time, its performance is impacted by the inherent complexity of short-window data.

\section{Discussion}\label{sec:discussion}

For the field to continue improving reading behavior classification, more ecological and in-the-wild research is required. Prior datasets and methods \cite{Chen2023, Kelton2019ReadingReal-time, Campbell2001} have not taken into consideration the impacts of classroom settings and the high-paced employment of distinct reading behaviors during authentic settings. We have illustrated how instructed and in-the-wild behaviors have different gaze measure distributions along critical variables such as velocity, density, and sequentiality. Through a combined qualitative and quantitative approach, we have developed a new exploratory theoretical framework along with a lightweight AI model to classify in-the-wild behaviors. While not claiming broad generalizability, this study lays the groundwork for understanding and analyzing naturalistic reading processes, underscoring the need for further research in this area.

\subsection{Instructed versus In-the-Wild Reader Behaviors}

Our Mann-Whitney $U$ test results have demonstrated the differences between instructed and in-the-wild behaviors for ``sequential'' and ``deep'' but not ``skimming''. This can be attributed to the inherent volatility of ``skim'' reading and/or how reading instruction can impact different reading behaviors unequally \cite{Kaakinen2010TaskReading}. Lastly, our recognition of new in-the-wild behaviors such as ``non-sequential'' highlights the gaps in understanding what other possible reading behaviors exist. Capturing the range of reader behaviors and their transitions during reading comprehension provides insights into how students navigate a text and the techniques they employ. It is these tactical shifts from ``sequential'' to ``deep'' or ``previewing/mapping'' that reveal the broader strategies students use, which would be overlooked if our learning theories and models did not account for these in-the-wild behaviors.

\subsection{Reading Behavior Framework and Taxonomy}

Our proposed framework and taxonomy for reading behaviors are based on human-coded observations of gaze scanpath videos, incorporating terms related to velocity, density, and sequentiality to distinguish between various behaviors. This approach is well-aligned with prior research on reading behaviors. For instance, in \citet{Carver1992ReadingImplications}, the concept of reading rate and "gear-shifting" is heavily reliant on WPM to differentiate between behaviors. Regarding density, \citet{Srivastava2018CombiningRecognition} found that the fixation dispersion area was the most useful feature in their gaze-based activity classifier, highlighting the importance of gaze spread in identifying reading behaviors. Lastly, the dimension of sequentiality builds on prior studies analyzing AOI sequences, such as those by \citet{Busjahn2015EyeOrder} and \citet{Ma2023FromReading}, which distinguish between novice and experienced readers by examining their different strategies for efficient information retrieval. By integrating these three dimensions and supporting literature, our framework provides comprehensive behavior archetypes with textual, visual, and statistical descriptions, enhancing the understanding of naturalistic reading processes.

\subsection{Real-Time Behavior Recognition}

The proposed 2D CNN model with a lightweight ResNet18 backbone achieved a macro F1 score of 0.8 under a challenging LOPOCV setup, demonstrating its capability for real-time reading behavior inference. Benchmarking experiments further support this, with the model achieving an average inference time of 3 ms on a GPU (RTX 3080) and 14 ms on a CPU (11th Gen Intel Core i7-11700F @ 2.50GHz), while gaze scanpath image generation takes 1.5 ms per frame. Given that the input frequency is based on a fixed 10-fixation window (~6 seconds), the processing pipeline operates significantly faster than the temporal resolution required for classification. The high variance in reading techniques during naturalistic reading presents a significant challenge for behavior identification, as readers employ a range of strategies from ``deep'' reading to ``skimming'', often switching behaviors based on context. This variance is particularly evident in the similarity between ``non-sequential'' and ``sequential'' reading, where subtle differences are harder to distinguish compared to more visually distinct behaviors like ``skimming''. The mixed ``instructed'' and ``in-the-wild'' methodology used in data collection helps capture this range of behaviors, but a larger and more diverse corpus would likely enhance model performance. Increasing the dataset size would allow the model to better generalize and adapt to the natural variability in reading techniques, ultimately improving its accuracy in real-world applications.

\section{Implications on Reading Instruction}

Real-time reading behavior recognition has the potential to revolutionize reading assessment and instruction by providing detailed, context-specific insights into how students interact with texts. By leveraging a classifier that can detect behaviors like ``skimming'', ``sequential reading'', or ``non-sequential reading'' in real-time, educators can access fine-grained data on how students engage with different sections of a passage. This allows for a more nuanced understanding of reading behaviors, such as identifying sections where a student skimmed through content, struggled, or read attentively. Such insights enable teachers to tailor their instruction based on specific behaviors observed during reading, offering targeted support to improve comprehension where needed.

Moreover, the ability to recognize reading behaviors in real-time with small context windows opens up the possibility of contextualized behavior mapping. For instance, if a student skims the first paragraph, reads the second paragraph regularly, and deeply reads the third, educators can quickly pinpoint where a student's attention was most and least focused. When linked to assessment questions related to specific text areas, this data becomes even more valuable. Educators can better understand if a student's incorrect answer resulted from skimming a relevant section or reading it without comprehension. This deeper insight into the reading process allows for more personalized and effective instructional interventions, enhancing overall literacy development by addressing specific reading challenges as they occur.

\section{Limitations and Future Work}

While this exploratory study provides valuable insights into reading behavior under different conditions, several limitations need to be acknowledged and addressed in future research. First, the study's small sample size of 27 participants may affect the generalizability of the findings. Future studies should consider increasing the number of participants to obtain more robust and widely applicable conclusions. Second, the use of a single PDF document as the reading material limited the variance in text layout and font size, potentially influencing reading behaviors in a constrained way. To enhance ecological validity, future research should explore a variety of reading materials with diverse layouts and content complexities, providing a more comprehensive understanding of reading behavior across different types of texts. Third, the study focused exclusively on a homogeneous participant population of 6th-grade students, which restricts the applicability of the findings to other age groups. Different age groups may exhibit distinct reading behaviors or employ varied strategies depending on the task. Future research should include a broader range of participants from different age groups and educational backgrounds to capture a more diverse range of reading behaviors. Lastly, the study only considered reading material with a single level of lexical complexity. It is hypothesized that varying levels of lexical complexity could significantly impact reading behaviors, with less experienced readers likely to struggle and slow down when engaging with more complex texts. Future studies should incorporate materials of varying lexical complexities to better understand how text difficulty influences reading behavior. By addressing these limitations, future research can build on the findings presented here to offer a more comprehensive understanding of reading behaviors in diverse and realistic contexts.

\section{Conclusion}\label{sec:conclusion}

This study provides a comprehensive exploration of reading behaviors in both instructed and naturalistic settings, leveraging a data-driven approach that integrates qualitative observations and quantitative data analysis to create a proof-of-concept framework for understanding and classifying reading behaviors. We examined how reading behaviors differ between these two contexts, finding that naturalistic settings reveal a greater range of behaviors, such as ``non-sequential'' and ``previewing/mapping'', which are not easily captured in instructed conditions. By developing a theoretical framework that encompasses reading behaviors and employing gaze metrics like WPM, inverse fixation dispersion, and FBSR, we effectively addressed \textbf{RQ1} by detailing the distinct characteristics and differences of reading behaviors in naturalistic versus instructed settings. For \textbf{RQ2}, we formulated a framework and taxonomy grounded in human observations that are measurable and distinguishable using data-driven methods, ensuring a comprehensive understanding of various reading behaviors. Finally, for \textbf{RQ3}, we demonstrated the efficacy of a 2D CNN model for classifying reading behaviors, achieving a Macro F1 score of 0.80, but also highlighted challenges with recognizing ``non-sequential'' behaviors due to class imbalance and data limitations, indicating the need for further research to refine AI-based behavior detection methods. 

This work advances our understanding of reading behaviors in real-world contexts by serving as an exploratory, proof-of-concept study. It aims to lay the groundwork for future research that can build upon these methods, develop more generalizable datasets, and refine models for broader applicability. By highlighting the feasibility of using AI to classify complex, in-the-wild reading behaviors, this study provides a critical first step toward more comprehensive and scalable approaches in educational technology and behavior recognition.

\begin{acks}

The research reported here was supported by the Institute of Education Sciences, U.S. Department of Education, through Grant R305A150199 and R305A210347 to Vanderbilt University. The opinions expressed are those of the authors and do not represent views of the Institute or the U.S. Department of Education. 

\end{acks}

\bibliographystyle{ACM-Reference-Format}
\bibliography{references, manual_references}

\appendix

\end{document}